\documentclass[aps,prc,superscriptaddress,showpacs,preprint]{revtex4-1}
\usepackage[english]{babel}
\usepackage{graphicx,amsmath,amsfonts,amssymb,amsbsy}
\usepackage[dvips]{color}
\usepackage{subfigure}
\usepackage{multirow}

\usepackage{pstricks}
\usepackage{mathrsfs}
\usepackage{lipsum}
\usepackage[normalem]{ulem}

\newcommand{\msun}{M_{\odot}}
\newcommand{\dblone}{\hbox{$1\hskip -1.2pt\vrule depth 0pt height 1.6ex width 0.7pt%
\vrule depth 0pt height 0.3pt width 0.12em$}}

\begin{document}

\author{M. Orsaria} \email{morsaria@rohan.sdsu.edu}
\affiliation{Department of Physics, San Diego State University, 5500
  Campanile Drive, San Diego, CA 92182, USA} \affiliation{CONICET,
  Rivadavia 1917, 1033 Buenos Aires, Argentina;\\ Gravitation,
  Astrophysics and Cosmology Group,\\ Facultad de Ciencias
  Astron{\'o}micas y Geof{\'i}sicas,
  UNLP,\\ Paseo del Bosque S/N (1900), La Plata, Argentina}

\author{H. Rodrigues}\email{harg@cefet-rj.br}
\affiliation{Centro
    Federal de Educa\c{c}\~ao Tecnol\'ogica do Rio de Janeiro, Av
    Maracan$\tilde{a}$ 249, 20271-110, Rio de Janeiro, RJ, Brazil}

\author{F. Weber} \email{fweber@mail.sdsu.edu}
\affiliation{Department of Physics, San Diego State University, 5500
    Campanile Drive, San Diego, California 92182}
\affiliation{Center for Astrophysics
    and Space Sciences, University of California,\\ San Diego, La Jolla,
    CA 92093, USA}

\author{G.~A.~Contrera} \email{guscontrera@gmail.com}
\affiliation{CONICET, Rivadavia 1917, 1033 Buenos Aires, Argentina}
\affiliation{IFLP, CONICET - Dpto. de F{\'i}sica, UNLP, La Plata, Argentina,}
\affiliation{Gravitation, Astrophysics
and Cosmology Group, Facultad de Ciencias Astron{\'o}micas y
    Geof{\'i}sicas, UNLP, Paseo del Bosque S/N (1900), La Plata, Argentina}

\title{Quark deconfinement in high-mass neutron stars}

\begin{abstract}
 In this paper, we explore whether or not quark deconfinement may
 occur in high-mass neutron stars such as J1614-2230 ($1.97 \pm 0.04\,
 \msun$) and J0348+0432 ($2.01 \pm 0.04 \, \msun$).  Our study is
 based on a non-local extension of the SU(3) Nambu Jona-Lasinio
 (n3NJL) model with repulsive vector interactions among the
 quarks. This model goes beyond the frequently used local version of
 the Nambu Jona-Lasinio (NJL) model by accounting for several key
 features of QCD which are not part of the local model. Confined
 hadronic matter is treated in the framework of non-linear
 relativistic mean field theory. We find that both the local as well
 as the non-local NJL model predict the existence of extended regions
 of mixed quark-hadron (quark-hybrid) matter in high-mass neutron
 stars with masses of 2.1 to $2.4\, \msun$.  Pure quark matter in the
 cores of neutron stars is obtained for certain parametrizations of
 the hadronic lagrangian and choices of the vector repulsion among
 quarks.  The radii of high-mass neutron stars with quark-hybrid
 matter and/or pure quark matter cores in their centers are found to
 lie in the canonical range of 12 to 13 km.
\end{abstract}




\maketitle

\section{Introduction}

White dwarf and neutron stars (NSs) are born in the aftermath of
core-collapsing supernova explosions.  Depending on NS mass and
rotational frequency, gravity may compress the matter in the core
regions of such stars up to more than ten times the density of
ordinary atomic nuclei, thus providing a high-pressure environment in
which numerous subatomic particle processes are likely to compete with
each other.  Theoretical studies indicate that hyperons, boson
condensates (pions, kaons, H-matter) and/or deconfined up, down and
strange quarks may exist in the core regions of NSs (for an overview,
see
\cite{glendenning00:book,fridolin1,blaschke01:trento,lattimer01:a,weber05:a,%
  page06:review,haensel06:book,klahn06:a_short,sedrakian07:a,klahn07:a,%
  alford08:a,CBMbook11:a} and references therein).

Based on qualitative considerations concerning the stiffness of the
nuclear equation of state (EoS), one could argue that the detection of
high-mass NSs, such as PSR J1614--2230 with a gravitational mass of
$1.97\pm 0.04\, M_{\odot}$ \cite{Demorest2010} and PSR J0348+0432 with
a mass of $2.01 \pm 0.04 \, \msun$ \cite{Antoniadis13}, may rule out
the existence of deconfined quarks in their central core regions,
since quark deconfinement would take away so much pressure that
high-mass NSs are not supported. As shown in Ref.\ \cite{Orsaria2013},
conclusions of that kind are not necessarily correct.

Our paper builds on the investigations carried out in
Ref.\ \cite{Orsaria2013} for NSs containing deconfined quark matter,
i.e.\ quark-hybrid stars (QHSs). The study is based on a generalized
version of the Nambu Jona-Lasinio (NJL) model \cite{Nambu}, which
accounts for several basic properties of quantum chromodynamics
(QCD).
As an improvement of the standard NJL model, we will consider
non-local interactions among the quarks \cite{Rip97}. By combining the
NJL model and the one-gluon-exchange model, which uses an effective
gluon propagator to model effective interactions among the quarks, it
is possible
\begin{table}[htb]
\begin{center}
\caption{Comparison of the key features of the local and non-local
  NJL model.}
\label{tabla5}
\begin{tabular}{ll}
\hline\hline
\hspace*{.1cm} Local NJL \hspace*{.1cm}  &\hspace*{.1cm} Non-local
NJL model \hspace*{.1cm} \\ \hline
                                  & $\bullet$ Confinement with a proper\\
  $\bullet$ Lack of confinement.   & choice of the non-local regulator\\
                                  & and model parameters \cite{Bowler:1994}. \\
\hline
   $\bullet$ Quark-quark scalar-         &  $\bullet$ Quark-quark interaction \\
   isoscalar and pseudoscalar-           & through phenomenologically        \\
   isovectorial local interaction.        & (effective) quark propagator.     \\
\hline
 $\bullet$ Non-renormalizable.           & $\bullet$ UV divergences are fixed \cite{Rip97}.\\
  Ultra-violet (UV) cutoff ($\Lambda$)   & Model dependent form factor $g(p)$.   \\
  is needed.              &        \\
\hline
  $\bullet$ Dynamical quark masses       & $\bullet$ Dynamical quark masses \\
  are momentum independent.              & are momentum dependent (as also \\
                                         & found in lattice QCD calculations)
\cite{Parappilly06}. \\
\hline
$\bullet$ Divergences in the meson        & $\bullet$ The momentum dependent \\
loop integrals. Extra cutoffs             & regulator makes the theory finite \\
are needed.                               & to all orders in the $1/N_c$ expansion
                                          \cite{Blaschke:1996}.\\
\hline
$\bullet$ The $\Lambda$ cutoff is turned off at high& $\bullet$ The
form factor provides a natural \\ momenta, limiting the applicability
& cutoff that falls off at high momenta. \\ of the model at high
densities. & \\ \hline \hline
\end{tabular}%
\end{center}
\end{table}
to introduce the non-locality in the quark-quark interaction
\cite{Blaschke1994} through a model-dependent form factor $g(p)$, in a
natural way. Table \ref{tabla5} compares the key features of the
standard (local) NJL with those of the non-local NJL model. Advantages
of the non-local model over the local model are indicated.

SU(2) versions of the nonlocal NJL model have been applied to the
study of hybrid stars in the past \cite{Dumm, Grunfeld:2007jt,
  g_contrera}. In this work, we model the quark phase that may exist
in the core of a neutron star using the non-local 3-flavor NJL (n3NJL)
model of Refs.\ \cite{Contrera2008} and \cite{Contrera2_2010}, which
includes vector interactions among the quarks. In our previous work
\cite{Orsaria2013}, we considered the vector interaction of the
non-local NJL model in a phenomenological way, and found that the
transition to pure quark matter occurs only in neutron stars which lie
already on the gravitationally unstable branch of the stellar
sequence. In this paper, we
explore the effect of the shift of the chemical potential on the form
factor, when vector interactions are considered. The results are
compared with those obtained by modeling the quark phase in the
framework of the local SU(3) NJL model (l3NJL) described, for example,
in Refs.\ \cite{hatsuda,rehberg,ratti,shao}.  The vector interactions
are know to be important for the QCD phase diagram \cite{kunih}. It is
therefore interesting, if not mandatory, to explore the consequences
of the vector interactions for the EoS of neutron star matter and the
structure of compact stars computed for such EoSs.

Global electric charge neutrality is imposed on the constituents of
neutron star matter. Local NJL studies carried out for local electric
charge neutrality have been reported recently in
Refs.\ \cite{Lenzi2012,Bonano,Masuda2012}.  In \cite{Lenzi2012} a
non-linear Walecka model was employed for the hadronic phase, using
parametrizations GM1 \cite{Glendenning1991} and NL3 \cite{Lalazissis},
and the local NJL model for the quark phase. In that work it was found
that the observation of compact stars with masses greater than around
$2\, M_{\odot}$ would be hard to explain unless one uses a very stiff
hadronic model for the nuclear EoS, such as NL3 with nucleons only,
instead of the softer GM1 EoS. The authors in reference \cite{Bonano}
treat vector interactions and color superconductivity among the quarks
in the framework of the local NJL model, using the NL3 and GM3
parametrizations for the description of confined hadronic matter. The
maximum mass of a neutron star was found to exceed $2 M_{\odot}$.  A
possible mixed phase of quarks and hadrons has not been considered in
\cite{Bonano}, whose appearance depends on the surface tension between
nuclear matter and quark matter, which is only very poorly known
\cite{glendenning00:book,Endo,glendenning92:a,glendenning01:a}. Screening and surface tension may be very important for understanding the quark-hadron phase transition and the existence of the mixed phase \cite{Endo}.

The authors in \cite{Providencia} analyze the possibility of quark matter
nucleation in high-mass neutron stars using the non-linear Walecka model plus the local NJL model with vector interaction for the EoS. They obtain stable NSs configurations with quark cores when the NL3 parametrization is considered (being $2.18M_{\odot}$ the largest NS mass) and NSs with mixed phase in their centers if the parametrization for the hadronic phase is TM1 or TM2 (being $2.03M_{\odot}$ the maximum star mass). 

Finally, we mention the study of Ref.\ \cite{Masuda2012}, where it was
found on the basis of the percolation picture from the hadronic phase
with hyperons to the quark phase with strangeness that massive neutron
stars with quark matter cores are compatible with the mass observed
for PSR J1614--2230, provided the crossover from hadronic matter to
quark matter takes place at around three times the normal nuclear
matter density, quark matter is strongly interacting in the crossover
region, and has a stiff equation of state.

This work is organized as follows. In Sect.\ \ref{sect:2}, we describe
the local as well as the non-local extension of the SU(3) NJL model at
zero temperature. In Sect.\ \ref{sect:3}, the non-linear relativistic
Walecka model, which is used to model confined hadronic matter, is
briefly discussed.  In Sect.\ \ref{sect:4}, we analyze the
construction of the mixed quark-hadron phase subjected to global
electric charge neutrality.  Our results for the quark-hadron
composition and bulk properties of neutron stars are presented in
Sects.\ \ref{sect:5} and \ref{sect:6}. Finally, a summary and
discussion of our results is provided in Sect.\ \ref{sect:7}.

\section{Quark Matter Phase}\label{sect:2}

\subsection{The local 3-flavor NJL model with vector interaction (l3NJL)}
\label{sect:2.1}

As and effective model of QCD, the NJL model accounts for the
interactions between constituent quarks and provides a simple scheme
for studying spontaneous chiral symmetry breaking, a key feature of
quantum chromodynamics (QCD) in the low temperature and density
domain, and its manifestations in hadron physics, such as dynamical
quark mass generation, the appearance of quark pair condensates, and
the role of pions as Goldstone bosons.  The effective action of the
local 3-flavor NJL model with vector interaction (l3NJL) used in this
paper is given by
\begin{widetext}
\begin{eqnarray}
S_E = \int d^4x \bigg\{&&\bar \psi (x)( i \partial{\hskip-2.0mm}/ -
{\hat m}) \psi (x) \;+\; \frac{1}{2}\,\,G_S [\,({\bar \psi
    (x)}\lambda_a \psi (x))^2 + ({\bar \psi (x)} i\gamma_5\lambda_a
  \psi (x))^2\,] \nonumber \\ &&+ H \,\left[ \,{\rm det} [{\bar \psi
      (x)}(1+\gamma_5) \psi (x)] + {\rm det} [{\bar \psi
      (x)}(1-\gamma_5) \psi (x)] \,\right]\; \nonumber \\ &&- G_V
[({\bar \psi (x)}\gamma^\mu\lambda_a \psi (x))^2 + ({\bar \psi (x)}
  i\gamma^\mu \gamma_5\lambda_a \psi (x))^2]\,\bigg\} ,
 \label{L3}
\end{eqnarray}
\end{widetext}
where $\psi$ is a chiral U(3) vector that includes the light quark
fields, $\psi \equiv (u, d, s)^T$, $\hat m = {\rm diag}(m_u, m_d,
m_s)$ is the current quark mass matrix, $\lambda_a$ with $a=1,...,8$
denote the generators of SU(3), and $\lambda_0=\sqrt{2/3}\,
\dblone_{3\times 3}$. The values for the coupling constants $G_S$ and
$H$ as well as the strange quark mass $m_{s}$ and the three-momentum
ultraviolet cutoff parameter, $\Lambda$, are model parameters. Their
values are taken from Ref.\ \cite{rehberg}, i.e., $m_u=m_d=5.5$ MeV,
$m_s=140.7$ MeV, $\Lambda=602.3$ MeV, $G_S\Lambda^2=3.67$ and
$H\Lambda^5=-12.36$. The vector coupling constant $G_V$ is treated as
a free parameter.

At the mean-field level, the thermodynamic potential associated with
$S_E$ is given by
\begin{widetext}
\begin{eqnarray}
\Omega^{L}(M_f, \mu) &=& G_S
\sum_{f=u,d,s}\left\langle\bar{\psi_{f}}\psi_{f}\right\rangle^2 + 4
H\left\langle\bar{\psi_{u}}\psi_{u}\right\rangle
\left\langle\bar{\psi_{d}}\psi_{d}\right\rangle\left\langle
\bar{\psi_{s}}\psi_{s}\right\rangle - 2
N_c\sum_{f=u,d,s}\int_\Lambda\frac{\mathrm{d}^3p}
{\left(2\pi\right)^3}\,{E_f} \nonumber \\ && -
\frac{N_c}{3\pi^2}\sum_{f=u,d,s}\int
_0^{p_{F_f}}\,\mathrm{d}{p}\,~\frac{p^4}{E_f} - G_V\,\sum_f
\rho_f^2\label{omega_njl} \, ,
\end{eqnarray}
\end{widetext}
where $N_c = 3$, $E_{f} = \sqrt{\mathbf{p}^{2}+M_{f}^{2}}$, and $p_{F_f} =
\sqrt{\mu_f^2-M_{f}^{2}}$.  The constituent quark masses $M_{f}$ are
given by
\begin{equation}
M_{f}=m_{f}-2G_S\left\langle\bar{\psi_{f}}\psi_{f}\right\rangle - 2H
\left\langle\bar{\psi_{j}} \psi_{j} \right\rangle\left\langle\bar
                {\psi_{k}} \psi_{k} \right\rangle \, ,
\end{equation}
with $f,j,k=u,d,s$ indicating cyclic permutations.  The vector
interaction shifts the quark chemical potential according to
\begin{equation}
\mu_f\, \rightarrow \mu_f - 2G_V\rho_f \, ,
\end{equation}
where $\rho_f$ is the quark number density corresponding to
  the flavor $f$ in the mean field approximation, that is,
\begin{equation}
\rho_f=\frac{N_c}{3\pi}[(\mu_f - 2G_V\rho_f)^2-M_f^2]^{3/2} \, .
\end{equation}
 The quark condensates $\left\langle\bar{\psi_{f}}\psi_{f}\right\rangle$ can be determined by minimizing the thermodynamic potential as
 $\left\langle\bar{\psi_{f}}\psi_{f}\right\rangle$, that is,
\begin{equation}
\label{loq}
\frac{\partial \Omega^{L}}{\partial
  \left\langle\bar{\psi_{f}}\psi_{f}\right\rangle}= 0\, , \quad
f=u,d,s \, .
\end{equation}

\subsection{The non-local 3-flavor model with vector interaction (n3NJL)}
\label{sect:2.2}

In this section we briefly describe the non-local extension of the
SU(3) Nambu Jona-Lasinio (n3NJL) model. The Euclidean effective action
for the quark sector, including the vector interaction, is given by
\begin{widetext}
\begin{eqnarray}
S_E &=& \int d^4x \left\{ \bar \psi (x) \left[ -i
  \partial{\hskip-2.0mm}/ + \hat m \right] \psi(x) - \frac{G_S}{2}
\left[ j_a^S(x) \ j_a^S(x) + j_a^P(x) \ j_a^P(x) \right] \right.
\nonumber \\ & & \qquad \qquad \left. - \frac{H}{4} \ T_{abc} \left[
  j_a^S(x) j_b^S(x) j_c^S(x) - 3\ j_a^S(x) j_b^P(x) j_c^P(x)
  \right]\right.
\left.- \frac{G_{V}}{2} \left[j_{V}^\mu(x)
j_{V}^\mu(x)\right]\right\}\;, \label{se}
\label{eq:S_E}
\end{eqnarray}
\end{widetext}
where $\psi$ and $\hat m$ stand for the light quark fields and the
current quark mass matrix, respectively.
For simplicity, we consider
the isospin symmetric limit in which case $m_u = m_d=\bar m$.  The
operator $\partial{\hskip-2.0mm}/ = \gamma_\mu\partial_\mu$ in
Euclidean space is defined as $\vec \gamma \cdot \vec \nabla +
\gamma_4\frac{\partial}{\partial \tau}$, with
$\gamma_4=i\gamma_0$. The Scalar, Pseudoscalar $j_a^{S,P}(x)$ and $j_{V}^{\mu}(x)$ vector currents
are, respectively, given by
\begin{align}
j_{a}^S(x) & =\int d^{4}z\ \widetilde{g}(z)\ \bar{\psi}\left(
x+\frac{z}{2}\right) \ \lambda_{a}\ \psi\left( x-\frac{z}{2}\right)
\ ,\nonumber\\ j_{a}^P(x) & =\int
d^{4}z\ \widetilde{g}(z)\ \bar{\psi}\left( x+\frac{z}{2}\right) \ i
\ \gamma_5 \lambda_{a} \ \psi\left(
x-\frac{z}{2}\right)\ ,\nonumber\\ j_{V}^{\mu}(x) & =\int
d^{4}z\ \widetilde{g}(z)\ \bar{\psi}\left( x+\frac{z}{2}\right)
\ \gamma^{\mu}\lambda_{a}\ \psi\left( x-\frac{z}{2}\right), \label{currents}
\end{align}
where $\widetilde{g}(z)$ is a form factor responsible for the
non-local character of the interaction and $\lambda_a$ represent the
generators of SU(3), as for the local model.

Finally, the constants $T_{abc}$ in the t'Hooft term, which account
for flavor-mixing, are defined by
\begin{equation}
T_{abc} = \frac{1}{3!} \ \epsilon_{ijk} \ \epsilon_{mnl} \
\left(\lambda_a\right)_{im} \left(\lambda_b\right)_{jn}
\left(\lambda_c\right)_{kl}\;.
\end{equation}

After standard bosonization of Eq.\ (\ref{se}), the integrals over the
quark fields can be evaluated in the framework of the Euclidean
four-momentum formalism. The thermodynamic potential, in the
mean-field approximation at zero temperature, can then be written as
\begin{widetext}
\begin{eqnarray}
\Omega^{NL} (M_f,0,\mu_f)  &=&  -\frac{N_c}{\pi^3}\sum_{f=u,d,s}
\int^{\infty}_{0} dp_0 \int^{\infty}_{0}\, dp \,\mbox{ ln }\left\{
\left[\widehat \omega_f^2 + M_{f}^2(\omega_f^2)\right]
\frac{1}{\omega_f^2 + m_{f}^2}\right\} \label{omzerot} \\ & & -
\frac{N_c}{\pi^2} \sum_{f=u,d,s} \int^{\sqrt{\mu_f^2-m_{f}^2}}_{0}
dp\,\, p^2\,\, \left[(\mu_f-E_f) \theta(\mu_f-m_f) \right]
\nonumber \\ & &
- \frac{1}{2}\left[\sum_{f=u,d,s} (\bar \sigma_f \ \bar S_f +
  \frac{G_S}{2} \ \bar S_f^2) + \frac{H}{2} \bar S_u\ \bar S_d\ \bar
  S_s\right]
- \sum_{f=u,d,s}\frac{\varpi_f^2}{4 G_V} \, , \nonumber
\end{eqnarray}
\end{widetext}
where $N_c=3$, $E_{f} = \sqrt{p^{2} + m_{f}^{2}}$, and $\omega_f^2 =
(\,p_0\,+\,i\,\mu_f\,)^2\,+\,p^2$.  The constituent quark masses
$M_{f}$ are treated as momentum-dependent quantities and are given by
\begin{equation}
M_{f}(\omega_{f}^2) \ = \ m_f + \bar\sigma_f g(\omega_{f}^2)\, ,
\end{equation}
where $g(\omega^2_f)$ is the Fourier transform of the form factor
$\widetilde{g}(z)$.  The inclusion of vector
interactions shifts the quark chemical potential as
\begin{equation}
\mu_f\, \rightarrow \widehat{\mu}_f=\mu_f - g(\omega^2_f)\varpi_f\, ,
\label{eq:mu_f}
\end{equation}
where $\varpi_f$ represent the vector mean fields related to the vector current interaction (last term in Eq.\ (\ref{eq:S_E})). The inclusion of the form factor in Eq.\ (\ref{eq:mu_f}) is a
particular feature of the non-local model, which renders the shifted
chemical potential momentum dependent.  Accordingly, the four momenta
$\omega_f$ in the dressed part of the thermodynamic potential are
modified as
\begin{equation}
\omega_f^2\,\rightarrow \widehat{\omega}_f^2 = (\,p_0\, +
\,i\,\widehat{\mu}_f\,)^2\, + \,p^2 \, .
\end{equation}
We followed the prescriptions given in \cite{Dumm} to include the
vector interaction.  Note that the quark chemical potential shift does
not affect the non-local form factor $g(\omega_{f}^2)$, as discussed
in \cite{Dumm,Weise2011,Contrera:2012wj}, avoiding a recursive
problem.

As we mentioned before, the form factor $\widetilde{g}(z)$ is
denfined by its Fourier transform in Euclidean space, which we take to be Gaussian
\begin{equation}
g(\omega^2_f) = \exp{\left(-\omega^2_f/\Lambda^2\right)}\, .
\label{form}
\end{equation}
It is worth noting that in Eq.\ref{form}, $\Lambda$ is not a cutoff as in the case of the local NJL, but a model parameter which plays a role for the width of the chiral
transition. This parameter as well as the quark current masses and
coupling constants in Eq.\ (\ref{se}) can be chosen so as to reproduce
the phenomenological values of pion decay constant $f_\pi$, and the
\begin{table}[htb]
\begin{center}
\caption{Parameters used for the non-local NJL (n3NJL)  model
  calculations presented in this paper.}
\label{tabla:n3NJL_param}
\begin{tabular}{cc}
\hline \hline
 Parameters~~~~~~~~  & n3NJL  \\
 \hline
$\bar m$   & $6.2$ MeV\\
$m_s$       & $140.7$ MeV \\
$\Lambda$   & $706.0$ MeV \\
$G_S\Lambda^2$   & $15.04$ \\
$H\Lambda^5$     & $-337.71$ \\
\hline \hline
\end{tabular}
\end{center}
\end{table}
meson masses $m_{\pi}$, $m_\eta$, $m_{\eta'}$, as described in
Refs.\ \cite{Contrera2008,Contrera2_2010}.  In this work we use for
the n3NJL model the parameters listed in Table \ref{tabla:n3NJL_param}
\cite{Orsaria2013}.

 Within the stationary phase approximation, the mean-field values of
 the auxiliary fields $\bar S_f$ turn out to be related to the
 mean-field values of the scalar fields $\bar \sigma_f$
 \cite{Scarpettini}. They are given by
\begin{equation}
\bar S_f = -\, 16\,N_c\, \int^{\infty}_{0}\,dp_0 \int^{\infty}_{0}
\frac{dp}{(2\pi)^3} \, g(\omega_f^2)\, \frac{
  M_{f}(\omega_f^2)}{\widehat{\omega}^2 + M_{f}^2(\omega_f^2)}\, .
\end{equation}

The mean field values of $\bar \sigma_u$, $\bar \sigma_s$ and
$\varpi_{f}$ are obtained via minimizing the thermodynamic potential,
\begin{equation}
\label{nonloq}
\frac{\partial \Omega^{NL}}{\partial \bar \sigma_f} = 0\, ,
\quad\frac{\partial \Omega^{\rm NL}}{\partial {\varpi_{f}}}= 0\, .
\end{equation}

\section{Hadronic Matter Phase}
\label{sect:3}

The hadronic phase is described in the framework of non-linear
relativistic field theory \cite{Walecka1974,Serot1986}, where baryons
(neutrons, protons, hyperons, delta states) interact via the exchange
of scalar, vector and isovector mesons ($\sigma $, $\omega$, $\rho $,
respectively). The parametrizations used in our study are GM1
\cite{Glendenning1991} and NL3 \cite{Lalazissis}. The associated
parameter values are summarized in Table \ref{tab:baryonic_couplings}.

The total lagrangian of the model is given by
\cite{glendenning00:book,fridolin1}
\begin{eqnarray}
\mathcal{L} = \mathcal{L}_{H} + \mathcal{L}_{\ell} \, ,
\label{eq:lag}
\end{eqnarray}
with the leptonic lagrangian given by
\begin{eqnarray}
\mathcal{L}_{\ell} = \sum_{\lambda=e^-, \mu^-} \bar{\psi}_\lambda
(i\gamma_\mu\partial^\mu - m_\lambda) \psi_\lambda \, .
\label{eq:leplag}
\end{eqnarray}
The hadronic lagrangian in Eq.\ (\ref{eq:lag}) has the form
\begin{widetext}
\begin{eqnarray}
  \mathcal{L}_{H} &=& \sum_{B=n,p, \Lambda, \Sigma, \Xi,
    \Delta}\bar{\psi}_B \bigl[\gamma_\mu (i\partial^\mu - g_\omega
    \omega^\mu - g_\rho \vec{\rho}_\mu) - (m_N - g_\sigma\sigma)
    \bigr] \psi_B  \nonumber \\
&&+ \frac{1}{2} (\partial_\mu \sigma\partial^\mu
  \sigma - m_\sigma^2 \sigma^2) - \frac{1}{3} b_\sigma m_N (g_\sigma
  \sigma)^3 - \frac{1}{4} c_\sigma (g_\sigma \sigma)^4 \nonumber \\
&&- \frac{1}{4}\omega_{\mu\nu} \omega^{\mu\nu}
  +\frac{1}{2}m_\omega^2\omega_\mu \omega^\mu + \frac{1}{2}m_\rho^2
  \vec{\rho}_\mu \cdot \vec{\rho\,}^\mu - \frac{1}{4}
  \vec{\rho}_{\mu\nu} \vec{\rho\,}^{\mu\nu} \, . \label{eq:Blag}
\end{eqnarray}
\end{widetext}
The quantity $B$ sums all baryonic particles which are produced in
neutron star matter at a given density
\cite{glendenning00:book,fridolin1}. Intriguingly (see
Sect.\ \ref{sect:6}), we find that, in addition to hyperons, the
$\Delta^-$ particle is generated in neutron star matter at densities
which are relevant for stable neutron stars. In contrast to this,
treatments of the quark-hadron phase transition based on the MIT bag
model \cite{glendenning92:a,glendenning01:a} do not predict the
occurrence of the $\Delta^-$ state.

The quantities $g_\rho$, $g_\sigma$, and $g_\omega$ in
Eq.\ (\ref{eq:Blag}) are meson-baryon coupling constants whose values
are summarized in Table \ref{tab:baryonic_couplings}.
\begin{table} [h]
  \caption{Parameters of the hadronic lagrangian of
    Eq.\ (\ref{eq:Blag}).  }
  \label{tab:baryonic_couplings}
\begin{ruledtabular}
\begin{tabular}{ccc}
   Coupling  &\multicolumn{2}{c}{Parametrizations}  \\
\cline{2-3}
 constants   &GM1    &NL3 \\
\hline
$g_\sigma$ & 8.910       &  10.217  \\
$g_\omega$ & 10.610      &  12.868 \\
$g_\rho$   & 8.196      &   8.948\\
$b_\sigma$ & 0.002947  &  0.002055 \\
$c_\sigma$ & $-0.001070$ & $-0.002651$  \\
\end{tabular}
\end{ruledtabular}
\end{table}

\begin{figure}[h]
\centering
\includegraphics[width=0.6 \textwidth]{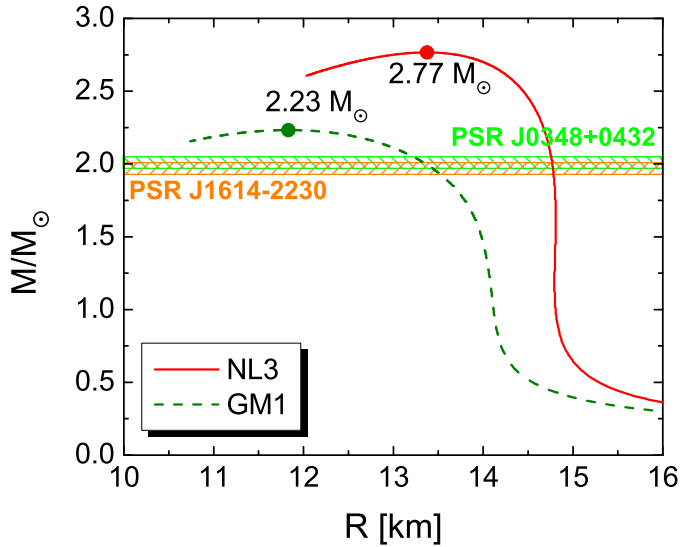}
\caption{(Color online) Mass-radius relationships of neutron stars
  computed for the pure (no quark matter) hadronic EoS studied in this
  work.}
\label{hadron}
\end{figure}

Table \ref{tab:saturation_properties} lists the properties of
symmetric nuclear matter computed from Eq.\ (\ref{eq:Blag}) for the
relativistic mean-field approximation.
\begin{table}
  \caption{Properties of symmetric nuclear matter at saturation
    density for the parameters listed in Table
    \ref{tab:baryonic_couplings}. Shown are the saturation density
    $\rho_0$, energy per baryon $E/N$, nuclear incompressibility $K$,
    effective nucleon mass $m^*_N$, and asymmetry energy $a_{\rm
      sy}$.  }
  \label{tab:saturation_properties}
\begin{ruledtabular}
\begin{tabular}{ccc}
              &\multicolumn{2}{c}{Parametrizations}  \\
\cline{1-3}
Properties                 &GM1    &NL3 \\
\hline
$\rho_0$ $({\rm fm}^{-3})$  &  0.153      &  0.148  \\
$E/N$ (MeV)               & $-16.3$      & $-16.3$  \\
$K$ (MeV)                 &300       &272  \\
$m^*/m_N$                 &  0.78         & 0.60  \\
$a_{\rm sy}$ (MeV)         &  32.5       &  37.4 \\
\end{tabular}
\end{ruledtabular}
\end{table}
The most important differences between the three parameter sets
concern the values of the nuclear incompressibility and the asymmetry
energy.  The maximum neutron star masses for GM1 and NL3 are $2.23\,
\msun$ and $2.77\, \msun$, respectively, for confined hadronic
matter. This is illustrated in Fig.\ \ref{hadron}, which shows the
mass-radius relationship of neutron stars for the parametrizations
(Table \ref{tab:baryonic_couplings}) studied in this work.

\section{Quark-Hadron Mixed Phase}
\label{sect:4}

The basic particle reactions in chemically equilibrated quark matter
are given by the strong process $u+d \leftrightarrow u+s$ and the weak
processes $d(s)\rightarrow u + e^{-}$ and $u+e^{-}\rightarrow d(s)$.
Neutrinos, once created by the weak reactions, do not accumulate in
cold neutron star matter, which implies zero chemical potentials for
(anti) neutrinos. The chemical potential for each quark flavor $f$ is
then given by
\begin{equation}
\mu_f = \mu_b - Q_f \, \mu_e \, ,
\end{equation}
where $\mu_b$ and $\mu_e$ denote the baryon and electron chemical
potential, respectively, and $Q_f$ stand for the electric charge of a
quark of flavor $f$. The baryon chemical potential is related to the
quark chemical potentials according to $\mu_b = 1/3 \sum_f \mu_f$.

The contribution of the leptons present in the quark matter phase to
the thermodynamic potential is given by
\begin{equation}
\label{elezero}
\Omega_{\lambda=e^{-} , \mu^{-}}(\mu_e) = - \frac{1}{\pi^2}
\int_{0}^{p_{F_{\lambda}}} p^2 \left(\sqrt{p^2 + m_{\lambda}^{2} } -
\mu_e \right) dp \, .
\end{equation}
Muons occur in the system if the electron chemical potential $\mu_e =
\mu_\mu$ is greater than the muon rest mass, $m_\mu = 105.7$~MeV. For
electrons we have $m_e = 0.511$ MeV. For the l3NJL model, the total
thermodynamic potential for the quark phase is then given by
Eqs.\ (\ref{omega_njl}) and (\ref{elezero}).  For the n3NJL model, the
total thermodynamic potential follows from Eqs.\ (\ref{omzerot}) and
(\ref{elezero}).

If the dense interior of a neutron star is indeed converted to quark
matter, it must be three-flavor quark matter since it has lower energy
than two-flavor quark matter. And just as for the hyperon content of
neutron stars, strangeness is not conserved on macroscopic time
scales, which allows neutron stars to convert confined hadronic matter
to three-flavor quark matter until equilibrium brings this process to
a halt.  As first realized by Glendenning
\cite{glendenning00:book,glendenning92:a,glendenning01:a}, the
presence of quark matter in neutron stars enables the hadronic regions
of the mixed phase to become more isospin symmetric than in the pure
phase by transferring electric charge to the quark phase. The symmetry
energy can be lowered thereby at only a small cost in rearranging the
quark Fermi surfaces. The electrons play only a minor role when
neutrality can be achieved among the baryon-charge carrying
particles. The stellar implication of this charge rearrangement is
that the mixed phase region of a neutron star will have positively
charged regions of nuclear matter and negatively charged regions of
quark matter
\cite{glendenning00:book,glendenning92:a,glendenning01:a}. This should
have important implications for the electric and thermal properties of
neutron stars.  First studies of the transport properties of
quark-hybrid neutron star matter have been reported in
\cite{reddy00:a,na12:a}.

To determine the mixed phase region of quarks and hadrons, we start
from the Gibbs condition for pressure equilibrium between confined
hadronic ($P^H$) matter and deconfined quark ($P^q$) matter.  The
Gibbs condition is given by
\cite{glendenning00:book,glendenning92:a,glendenning01:a}
\begin{eqnarray}
 P^H(\mu_b^H, \mu_e^H, \{\phi \} ) = P^q(\mu_b^q, \mu_e^q, \{\psi \} )
 \, ,
\label{eq:GibbsP}
\end{eqnarray}
with $\mu_b^H = \mu_b^q$ for the baryon chemical potentials and
$\mu_e^H=\mu_e^q$ for the electron chemical potentials in the hadronic
($H$) and quark ($q$) phase, respectively. By definition, the quark
chemical potential is given by $\mu_b^q=\mu_n/3$, where $\mu_n$ is the
chemical potential of the neutron. The quantities $\{ \phi \}$ and $\{
\psi\}$ in Eq.\ (\ref{eq:GibbsP}) stand collectively for the field
variables and Fermi momenta that characterize the solutions to the
equations of confined hadronic matter and deconfined quark matter,
respectively.  In the mixed phase, the baryon number density, $n_b$,
and the energy density, $\varepsilon$, are given by
\cite{glendenning00:book,glendenning92:a,glendenning01:a}
\begin{equation}
    n_b = (1-\chi) n_b^H + \chi n_b^q \, ,
\end{equation}
and
\begin{equation}
    \varepsilon = (1-\chi) \varepsilon^H + \chi\varepsilon^q\, ,
\end{equation}
where $n_b^H$ ($\varepsilon^H$) and $n_b^q$ ($\varepsilon^q$) denote
the baryon number (energy) densities of the hadron and quark phase,
respectively. The quantity $\chi \equiv V_q/V$ denotes the volume
proportion of quark matter, $V_q$, in the unknown volume $V$. By
definition, $\chi$ therefore varies between 0 and 1, depending on how
much confined hadronic matter has been converted to quark matter
\cite{glendenning00:book,glendenning92:a,glendenning01:a}. In
addition to the Gibbs condition (\ref{eq:GibbsP}) for pressure, the
conditions of global baryon number conservation and global electric charge
neutrality need to be imposed on the field equations.  The global
conservation of baryon charge is expressed as
\cite{glendenning00:book,glendenning92:a,glendenning01:a}
\begin{eqnarray}
  \rho_b = \chi \, \rho_Q(\mu_n, \mu_e ) + (1-\chi) \, \rho_H (\mu_n,
  \mu_e, \{ \phi \}) \, ,
\label{eq:mixed_rho}
\end{eqnarray}
where  $\rho_Q$ and $\rho_H$ denote the baryon number densities of the
quark phase and hadronic phase, respectively. The condition of global
electric charge neutrality is given by the equation
\begin{equation}
(1 - \chi) \sum_{i=B,l} q_i^H \, n_i^H + \chi \sum_{i=q,l}q_i^q \,
  n_i^q = 0 \, ,
\end{equation}
where $q_i$ is the electric charge of the $i$-th specie in units of
the electron charge.  In this work we have chosen global rather than
local electric charge neutrality. The latter is not fully consistent
with the Einstein-Maxwell equations and the micro physical condition
of chemical equilibrium and relativistic quantum statistics, as shown
in \cite{Rufini2011}. In contrast to local electric charge neutrality,
the global neutrality condition puts a net positive electric charge on
confined hadronic matter, rendering it more isospin symmetric, and a
net negative electric charge on the deconfined quark phase, allowing
neutron star matter to settle down in a lower energy state that
otherwise possible \cite{glendenning92:a,glendenning01:a}.

\section{Models for the ultra-dense part of the EoS of  neutron star matter}
\label{sect:5}

\begin{figure}[h]
\centering \includegraphics[width=0.6\textwidth]{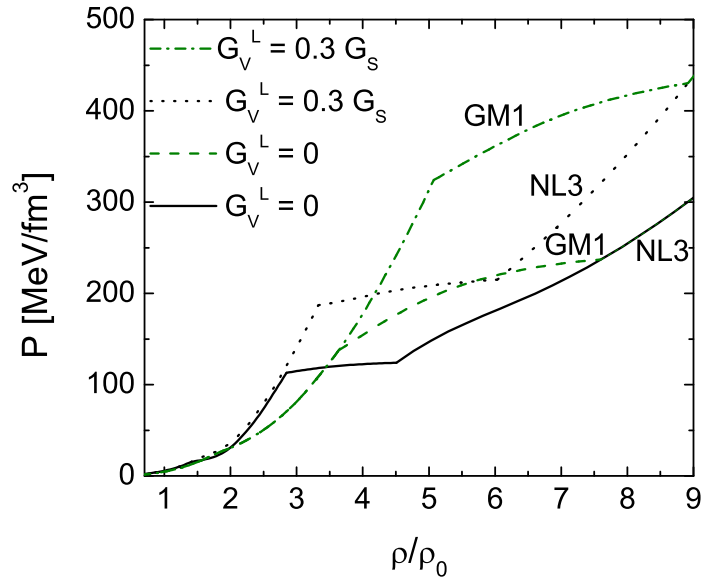}
\caption{(Color online) Pressure as a function of baryon number
  density for the local NJL model, l3NJL. The hadronic
  parametrizations are GM1 and NL3, and the vector repulsion strengths
  are $G_V^L/G_S =0$ and $G_V^L/G_S=0.3$.}
  \label{fig:eos1}
\end{figure}

\begin{figure}[!htbp]
\centering
\includegraphics[width=0.6 \textwidth]{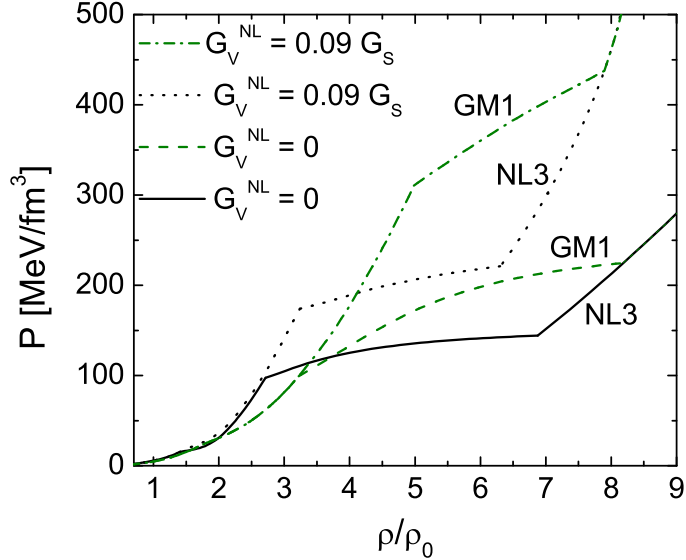}
\caption{ (Color online) Same as Fig.\ \ref{fig:eos1}, but for the
  non-local NJL model, n3NJL, and vector repulsion strengths
  $G_V^{NL}/G_S =0$ and $G_V^{NL}/G_S=0.09$.}
\label{fig:eos2}
\end{figure}

\begin{figure}[!htbp]
\centering
\includegraphics[width=0.6 \textwidth]{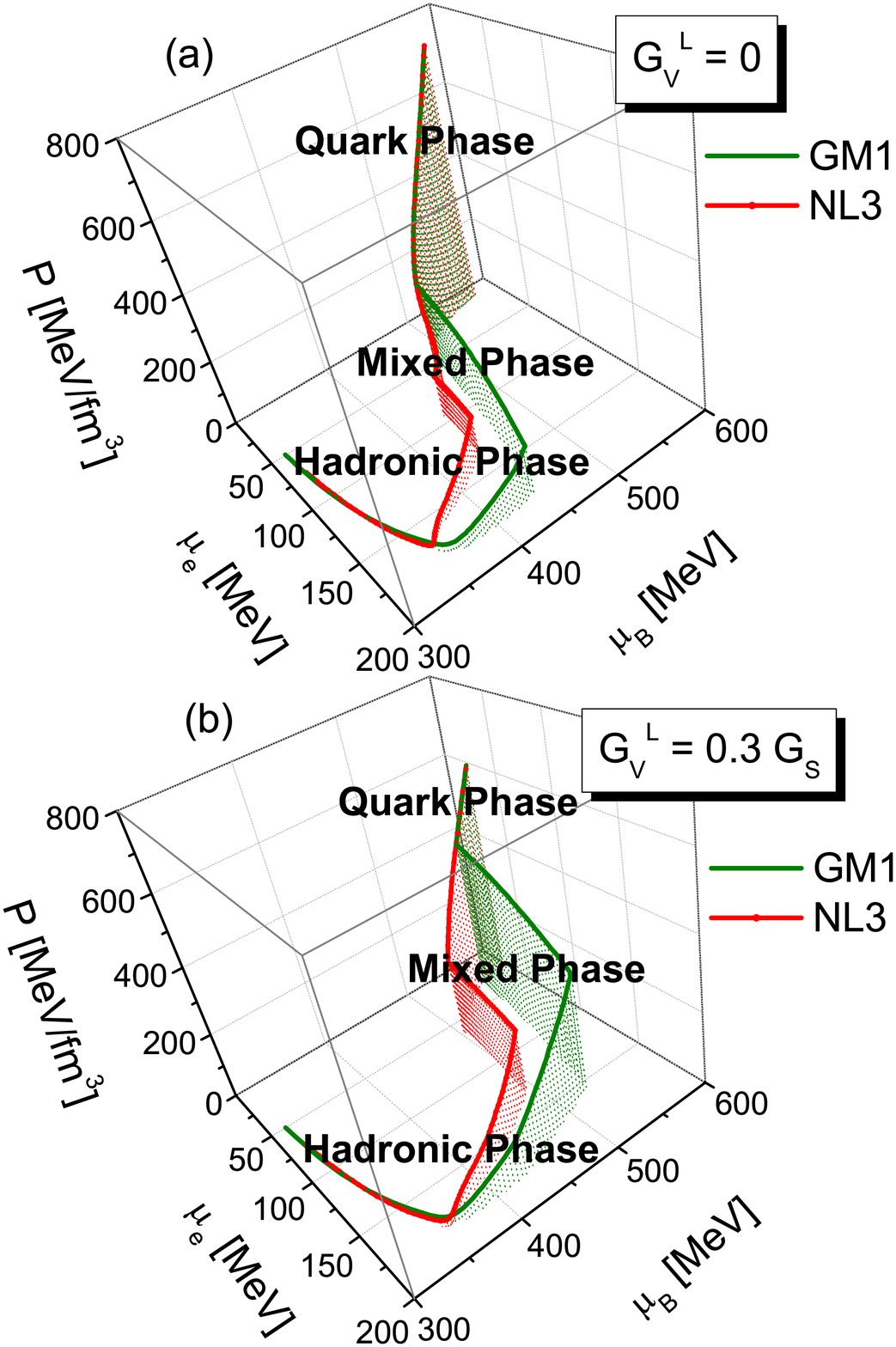}
\caption{(Color online) Pressure as a function of baryon
  ($\mu_B = \mu_n/3$)
and electron ($\mu_e$) chemical potential for
  the local NJL model, l3NJL, hadronic parametrizations GM1 and NL3,
  and vector repulsions (a) $G_V^{L}/G_S =0$ and (b) $G_V^{L}/G_S =0.3$.}
\label{fig:eos_3D0}
\end{figure}

\begin{figure}[!htbp]
\centering
\includegraphics[width=0.6 \textwidth]{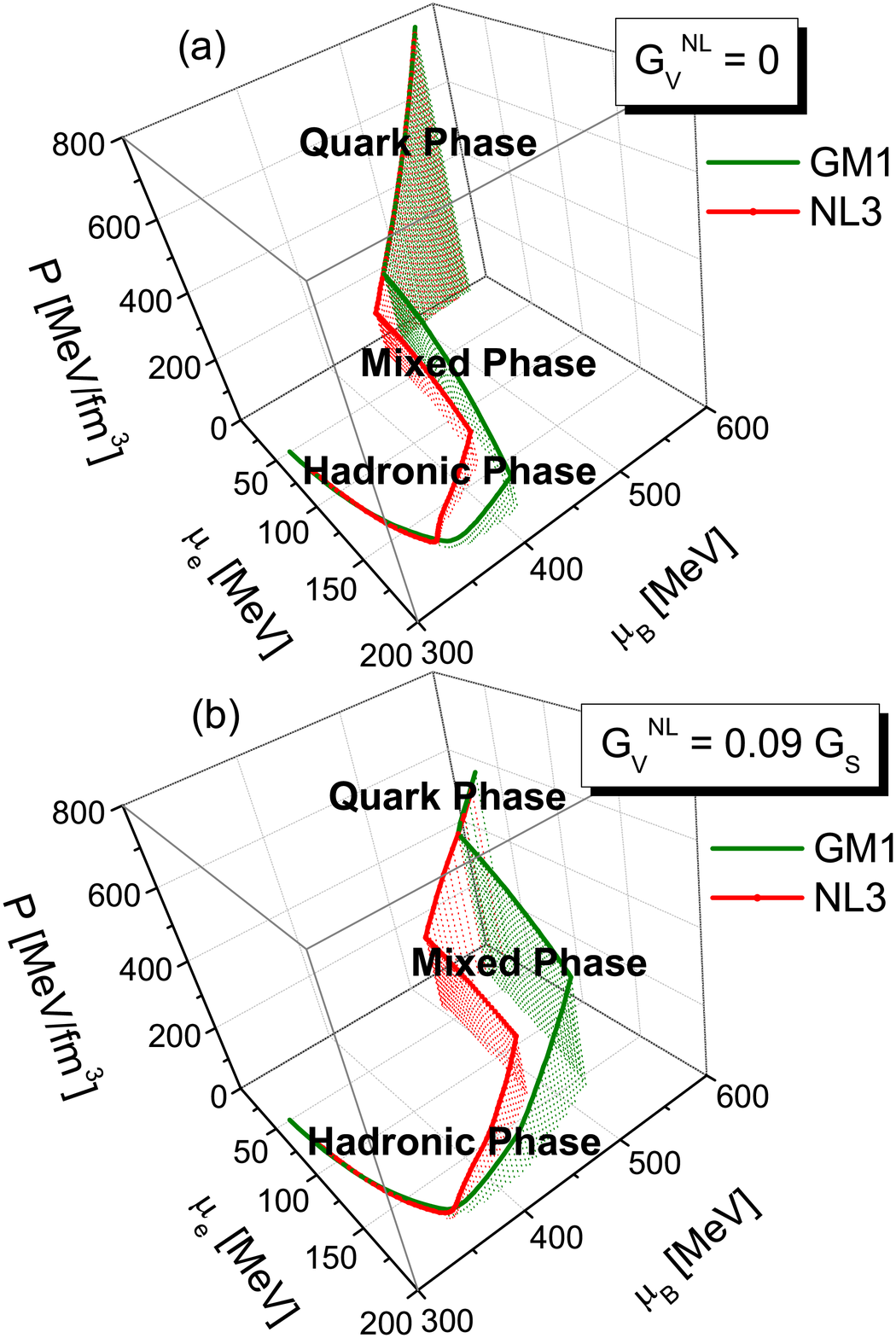}
\caption{(Color online) Same as Fig.\ \ref{fig:eos_3D0}, but for
  the non-local NJL model, n3NJL, hadronic parametrizations GM1 and
  NL3, and vector repulsions (a) $G_V^{NL}/G_S =0$ and (b)
  $G_V^{NL}/G_S=0.09$.}
\label{fig:eos_3D}
\end{figure}

 Figures \ref{fig:eos1} and \ref{fig:eos2} show the EoS of neutron
 star matter computed for the local (l3NJL) and non-local (n3NJL)
 model, respectively. The hadronic contributions are computed for the
 lagrangian given in Eq.\ (\ref{eq:lag}). For the quark matter phase
 the local NJL model (Sec.\ \ref{sect:2.1}) and the non-local NJL
 model (Sec.\ \ref{sect:2.2})  have been used.
 For the hadronic phase we consider the parameter sets GM1 and NL3.
 The quark phase is being investigated for repulsive vector
 interactions $G_V$ among quarks which range from zero to the upper
 limits set by the local and non-local model.
The equations of state shown in Figs.\ \ref{fig:eos1} and
\ref{fig:eos2} are plotted in the three-space spanned by the neutron
chemical potential, electron chemical potential and pressure in
Figs.\ \ref{fig:eos_3D0} and \ref{fig:eos_3D}.

\begin{figure}[!htbp]
\centering
\includegraphics[width=0.65 \textwidth]{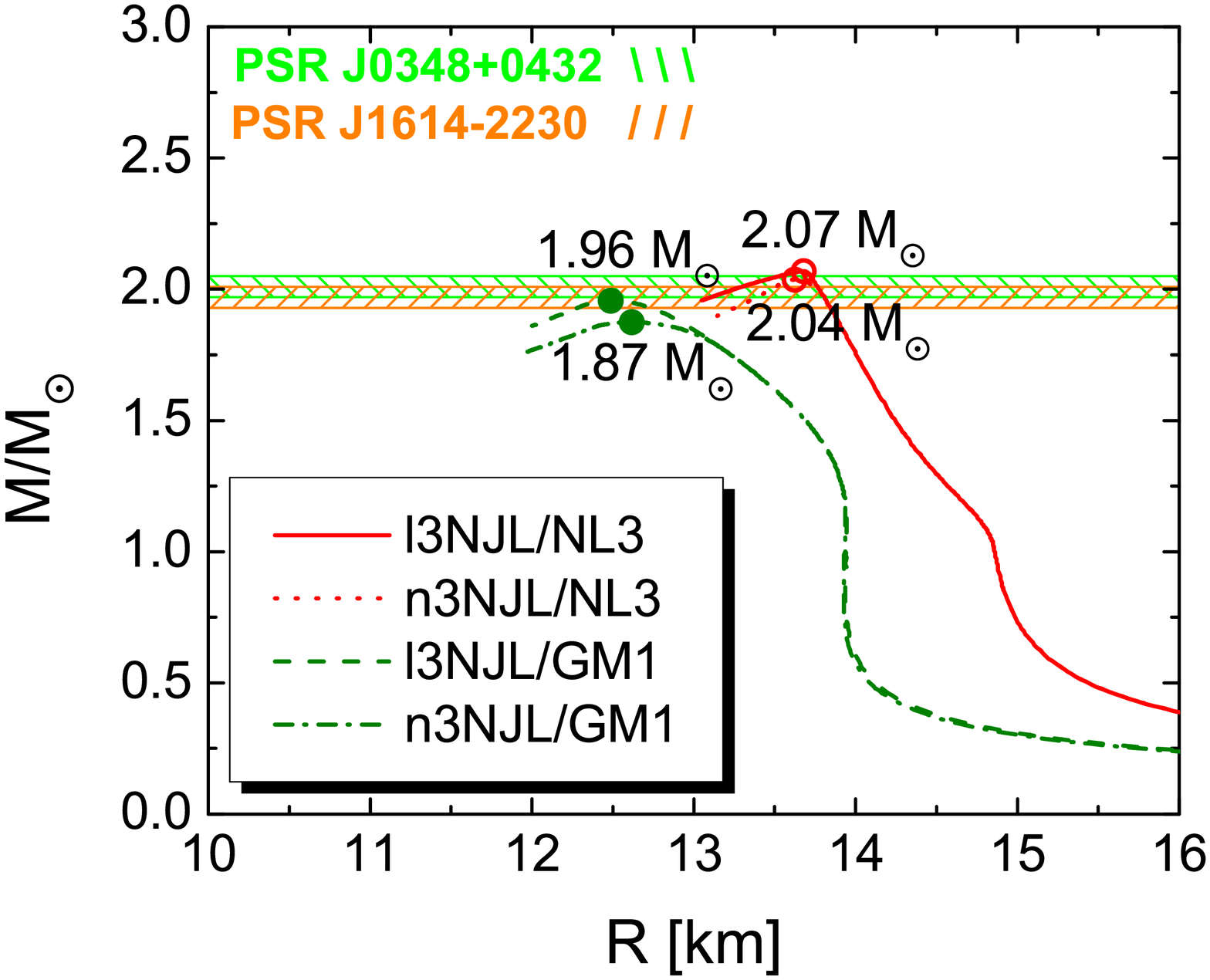}
\caption{(Color online) Same as Fig.\ \ref{hadron}, but with quark
  matter included. The vector repulsion is $G_V =0$.}
\label{hadron2}
\end{figure}

Fig.\ \ref{hadron2} shows the mass-radius relationship of neutron
stars for the three selected parametrizations of the hadronic
lagrangian of this work.
 The transition to quark matter is included, but the vector repulsion
 among quarks is switched off. This lowers the maximum masses of the
 neutron stars (compare with Fig.\ \ref{hadron}), since a standard
 treatment of the quark-hadron transition softens the equation of
 state. The masses of the recently discovered high-mass neutron stars
 J1614-2230 ($1.97 \pm 0.04\, \msun$) \cite{Demorest2010} and
 J0348+0432 ($2.01 \pm 0.04 \, \msun$) \cite{Antoniadis13} are shown
 for comparison. Our calculations show that even for zero vector
 repulsion both stars could contain quark-hybrid matter in their
 cores.

\section{Quark-Hybrid Composition}
\label{sect:6}

\begin{figure}[!htbp]
\centering
\includegraphics[width=0.6 \textwidth]{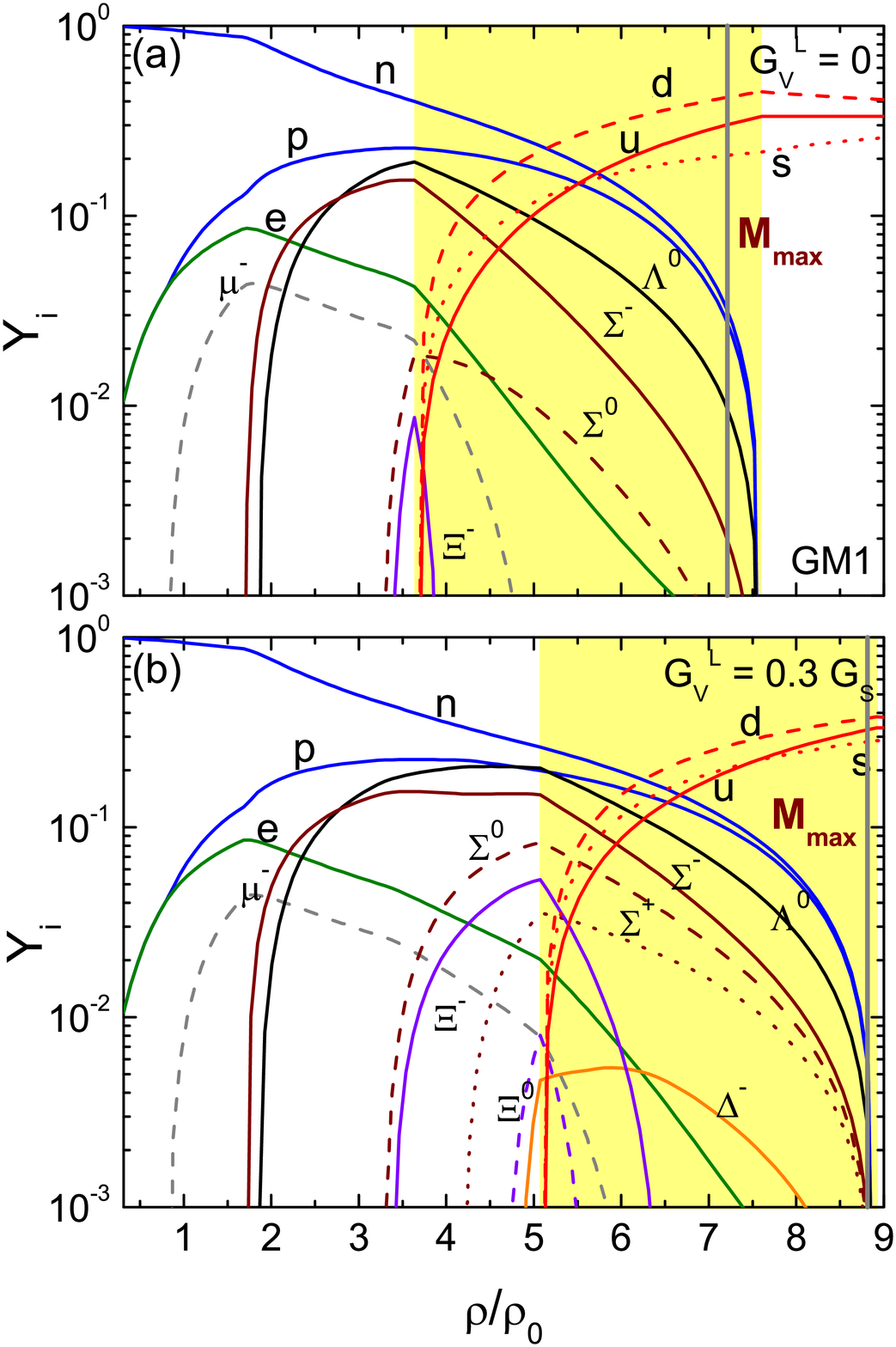}
\caption{(Color online) Particle population of neutron star matter
  computed for the local NJL model, l3NJL. Yellow areas highlight the
  mixed phases. The solid vertical lines indicate the central
  densities of the associated maximum-mass NSs. The hadronic model
  parametrization is GM1 and the vector repulsions are (a)
  $G_V^{L}/G_S =0$ and (b) $G_V^{L}/G_S =0.3$.}
\label{pb0}
\end{figure}

\begin{figure}[!htbp]
\centering
\includegraphics[width=0.6 \textwidth]{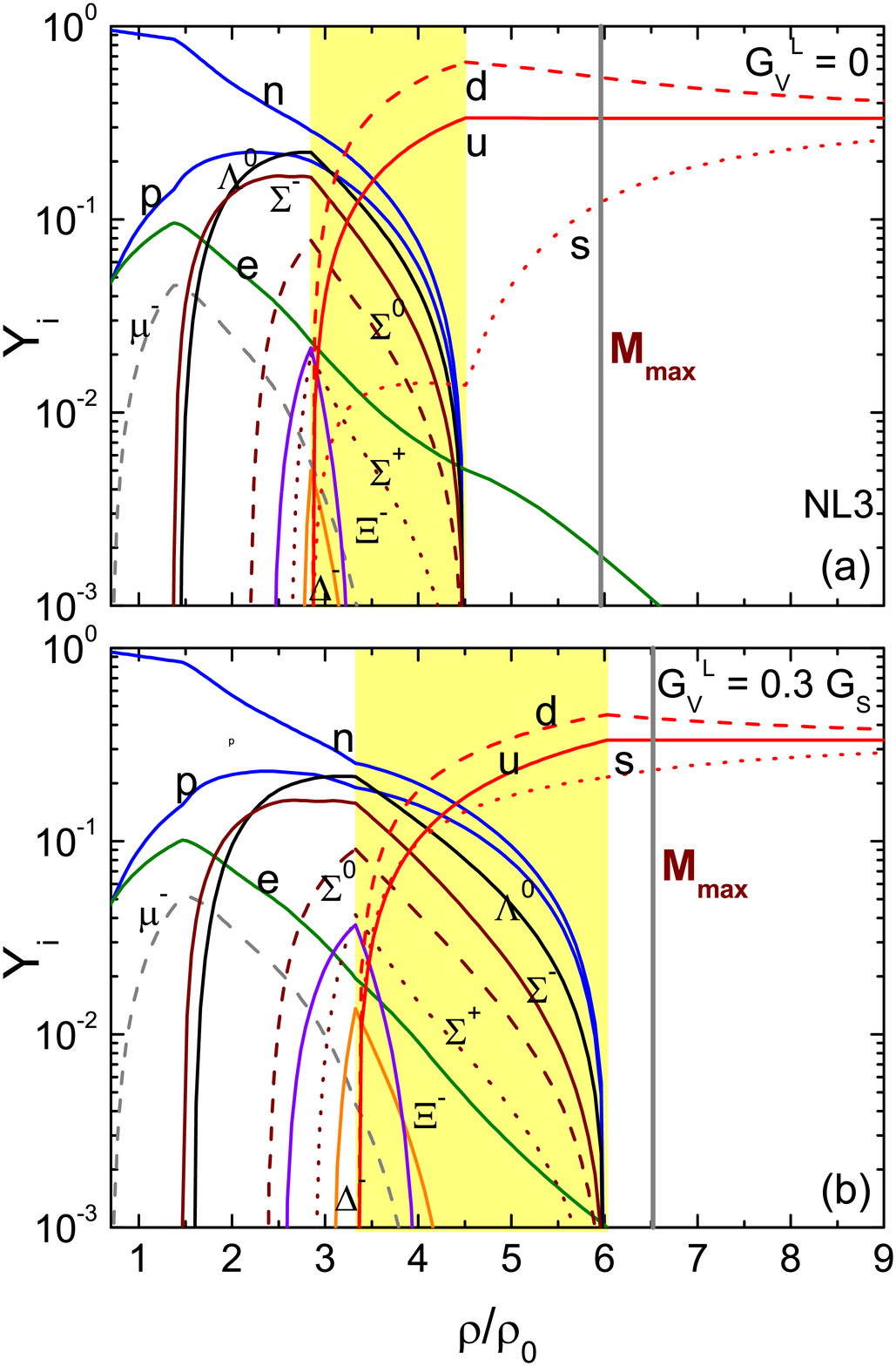}
\caption{(Color online) Same as Fig.\ \ref{pb0}, but for the hadronic
  model parametrization NL3 and vector repulsions (a) $G_V^{L}/G_S =0$
  and (b) $G_V^{L}/G_S =0.3$.  \label{pb0203}}
\end{figure}

\begin{figure}[!htbp]
\includegraphics[width=0.6 \textwidth]{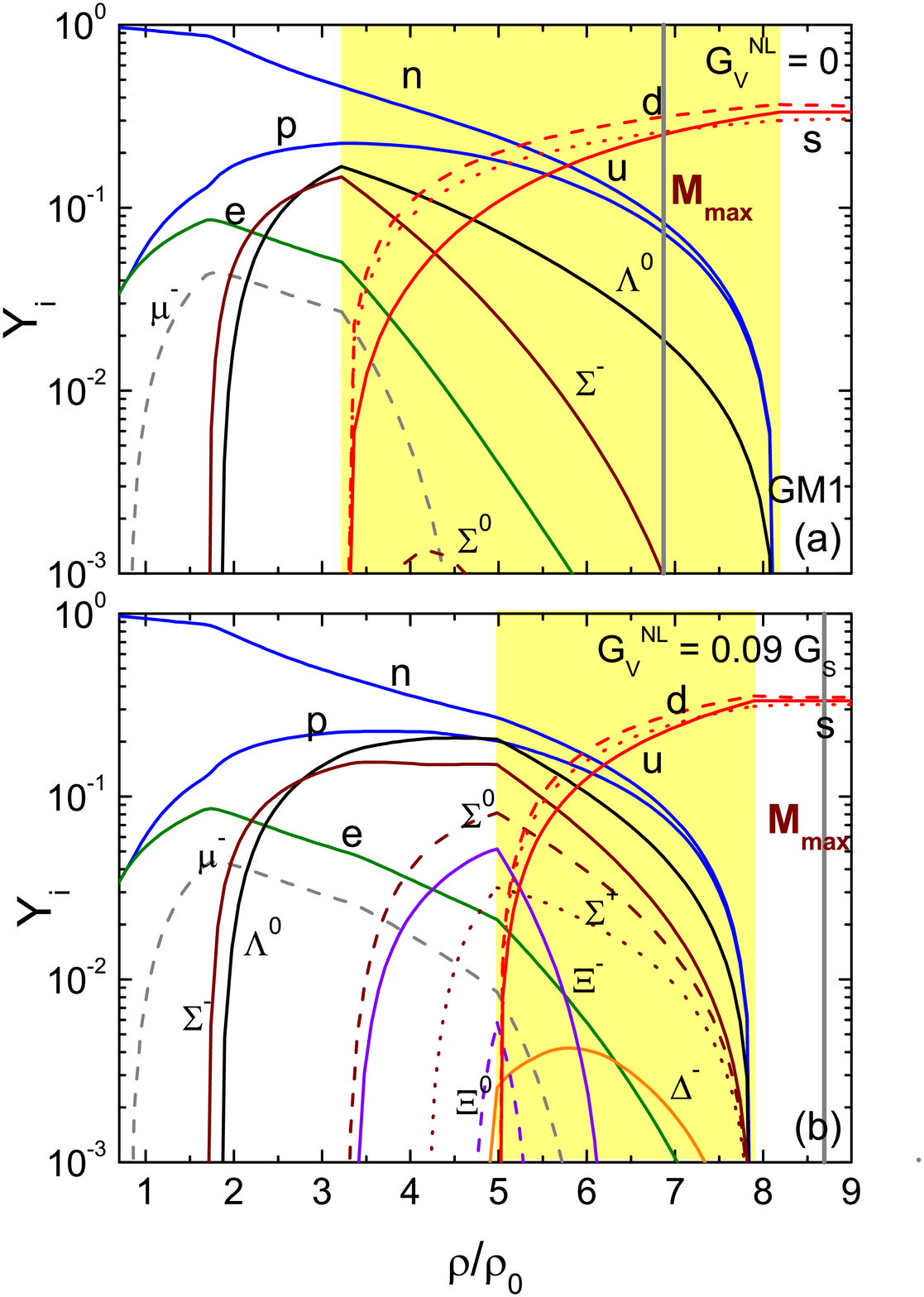}
\caption{(Color online) Particle population of neutron star matter
  computed for the non-local NJL model, n3NJL. The yellow areas
  highlight the mixed phase. The solid vertical lines indicate the
  central densities of the associated maximum-mass NSs. The hadronic
  model parametrization is GM1 and the vector repulsions are (a)
  $G_V^{NL}/G_S =0$ and (b) $G_V^{NL}/G_S =0.09$.}
\label{pb005}
\end{figure}

\begin{figure}[!htbp]
\includegraphics[width=0.6 \textwidth]{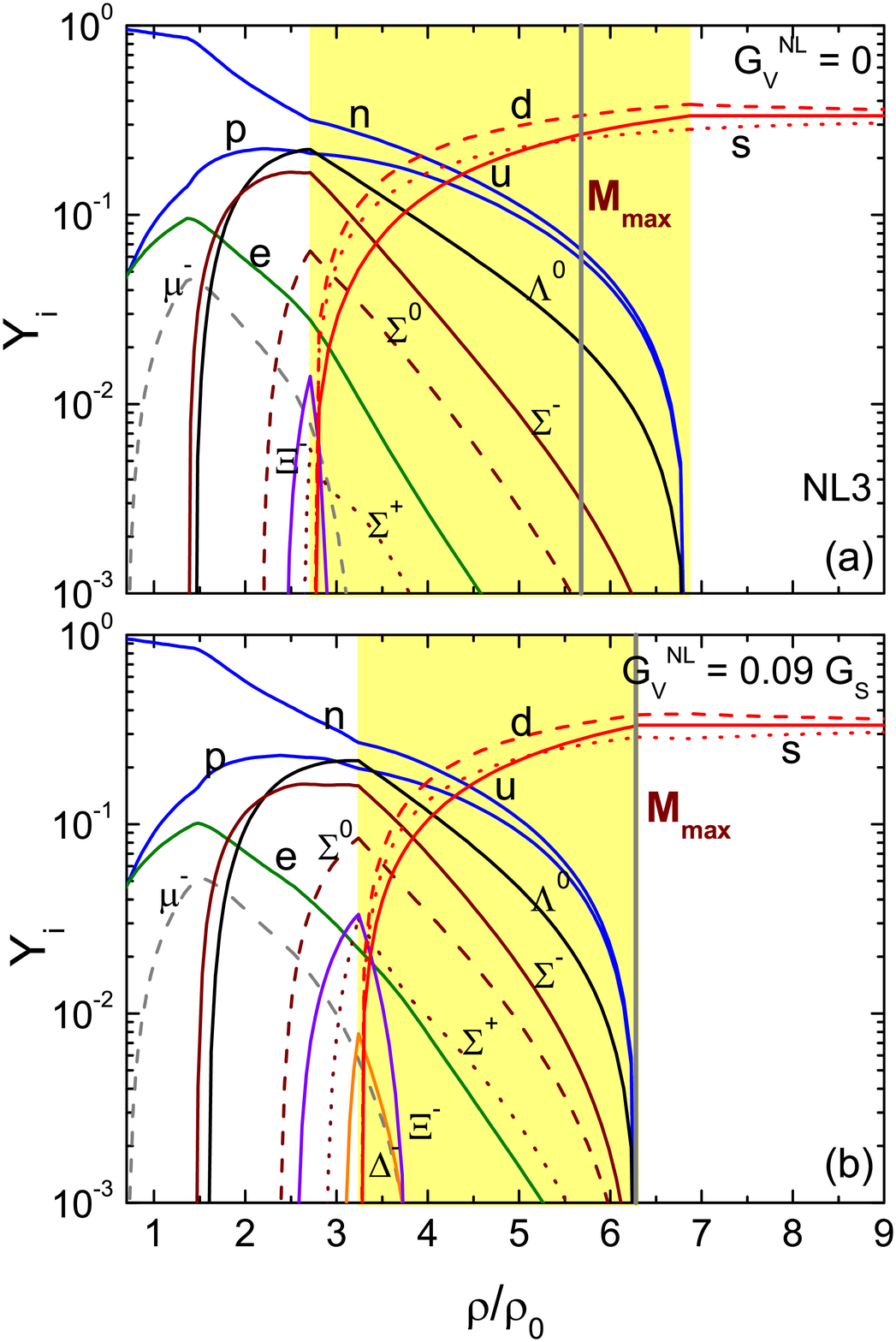}
\caption{(Color online) Same as Fig.\ \ref{pb005}, but for the
  hadronic model parametrization NL3 and vector repulsions (a)
  $G_V^{NL}/G_S =0$ and (b) $G_V^{NL}/G_S =0.09$.}
\label{pb}
\end{figure}

In Figs.\ \ref{pb0} through \ref{pb} we show the relative particle
fractions $Y_i$ $(\equiv \rho_i/\rho)$ of neutron star mater as
a function of baryon number density for both the local and non-local
NJL model. It can be seem that by increasing the strength of the
vector interaction, negatively charged particles like $\mu^{-}$'s and
$\Delta^{-}$'s take on the role of electrons, whose primary duty is to
make the stellar matter
electrically neutral. Because of the early onset of the $\Delta$
population in our models, there is less need for electrons so that their
number density in the mixed phase is reduced compared to the
\begin{table}[htb]
\begin{center}
\caption{Deleptonization densities for the local and non-local NJL
  model, for the GM1 parametrization ($\rho_0 = 0.16 \, {\rm
    fm}^{-3}$).}
\label{tabla6}
\begin{ruledtabular}
\begin{tabular}{ccc}
          &\multicolumn{2}{c}{Deleptonization density ($\rho_0$)}\\
\cline{2-3}
$G_V / G_S$   &Local NJL       &Non-local NJL \\
\hline
$0$ (GM1)    &  $6.65$  &$5.85$  \\
$0$ (NL3)    &  $6.61$  &$4.55$  \\
$0.09$ (GM1) &  $-$  &$7.03$  \\
$0.09$ (NL3) &  $-$  &$5.33$     \\
$0.30$ (GM1) &  $7.45$  &$-$     \\
$0.30$ (NL3) &  $6.26$  &$-$     \\
\end{tabular}
\end{ruledtabular}
\end{center}
\end{table}
outcome of standard mean-field/bag model calculations.  Table
\ref{tabla6} shows the densities beyond which leptons are no
  longer present in quark-hybrid matter (quark matter gets
  deleptonized). They are different for the local and non-local NJL
model and depend on the ratio $G_V/G_S$.

\begin{figure}[!htbp]
\centering
\includegraphics[width=0.6 \textwidth]{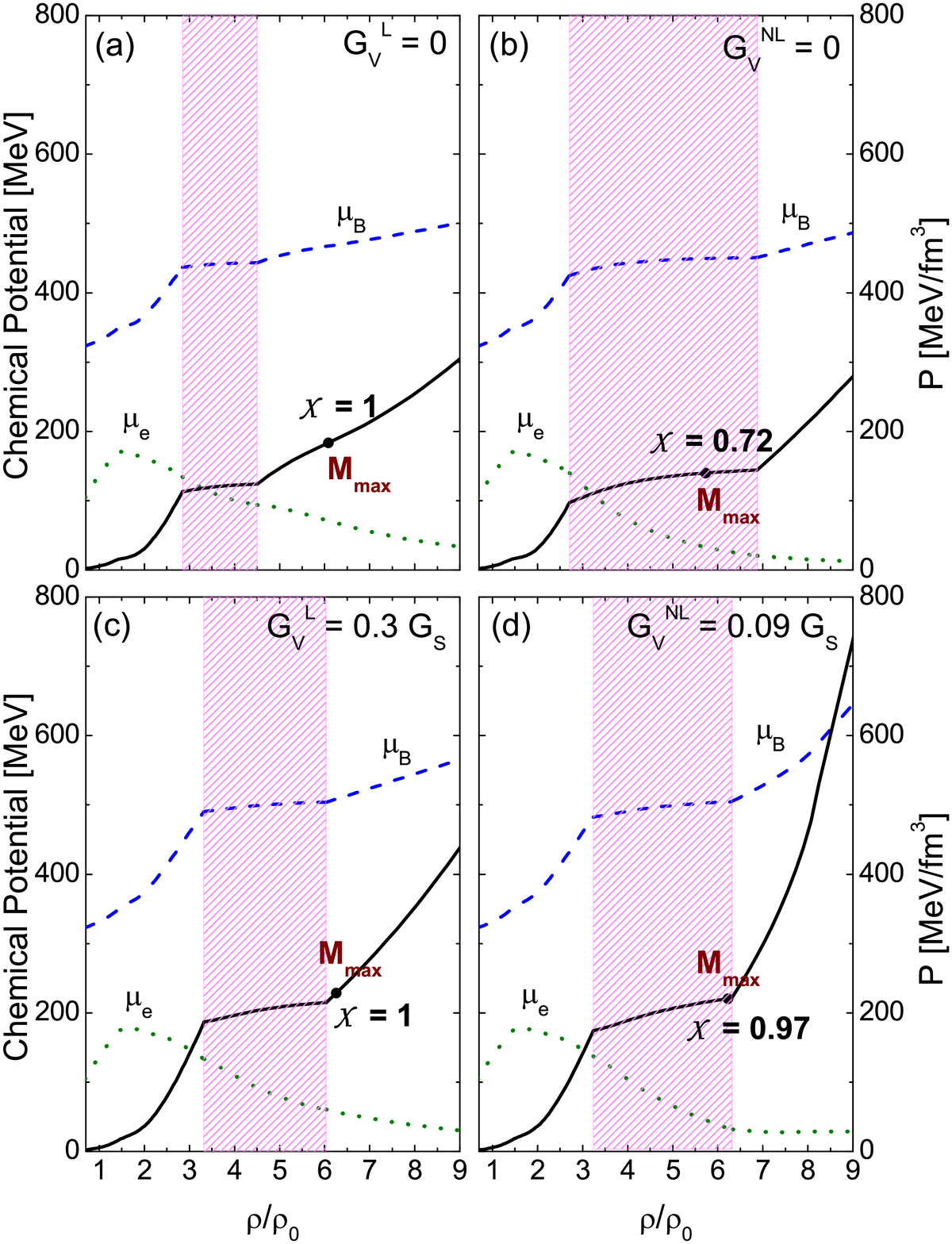}
\caption{(Color online) Pressure $P$ (solid lines), baryon chemical
  potential $\mu_B = \mu_n/3$ (dashed lines), and electron chemical
  potential $\mu_e$ (dotted lines) as a function of baryon number
  density (in units of $\rho_0 = 0.16 \, {\rm fm}^{-3}$) for
  parametrization NL3. The hatched areas denote the mixed phase
  regions where confined hadronic matter and deconfined quark matter
  coexist. Panels (a) and (b) are computed for l3NJL and n3NJL,
  respectively, and zero vector repulsion. The impact of finite values
  of the vector repulsion ($0.3 G_S$ and $0.09 G_s$) on the data is
  shown in panels (c) and (d) for l3NJL and n3NJL, respectively.}
\label{press}
\end{figure}

\begin{figure}[!htbp]
\centering
\includegraphics[width=0.6\textwidth]{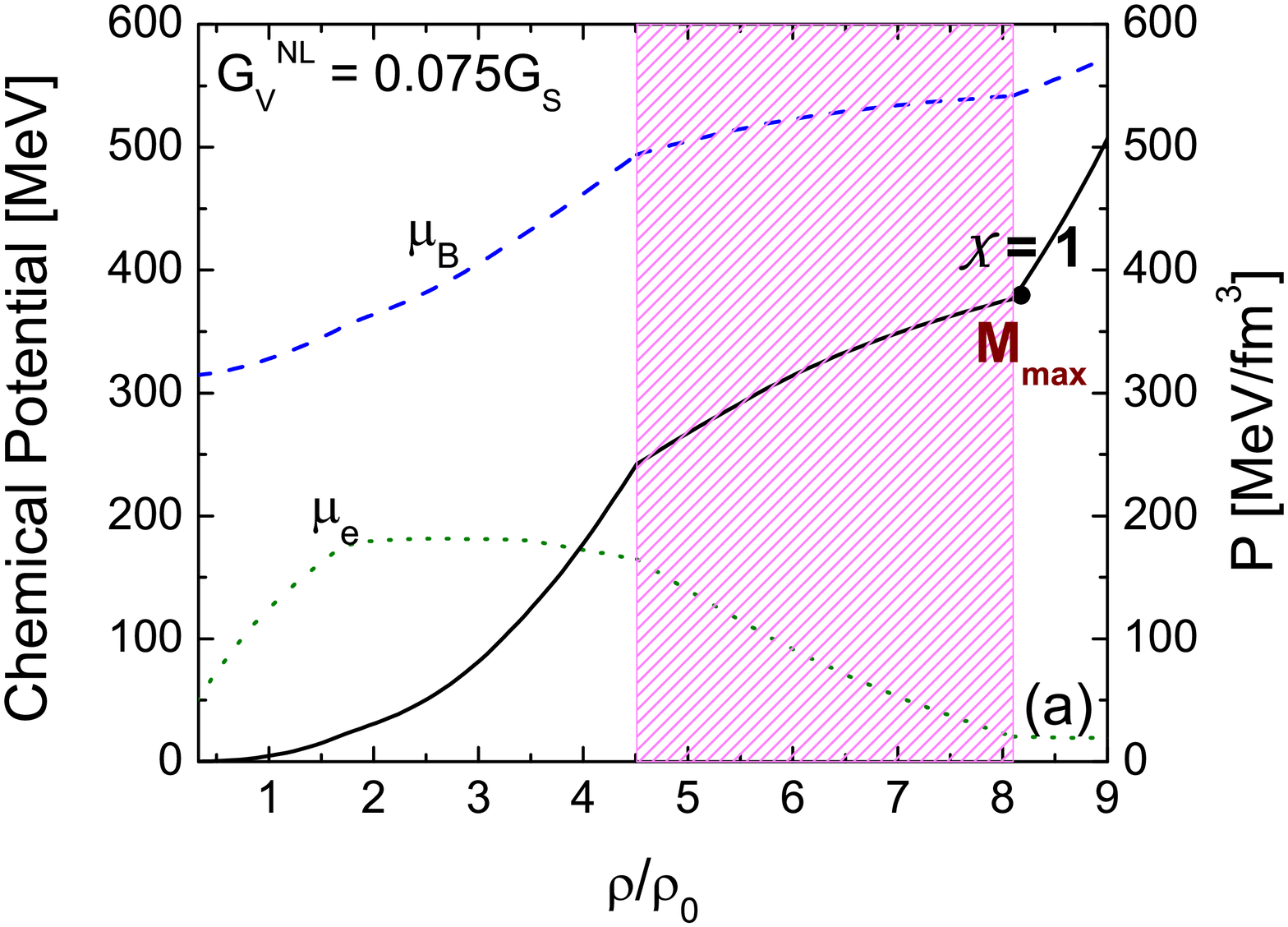}
\includegraphics[width=0.6\textwidth]{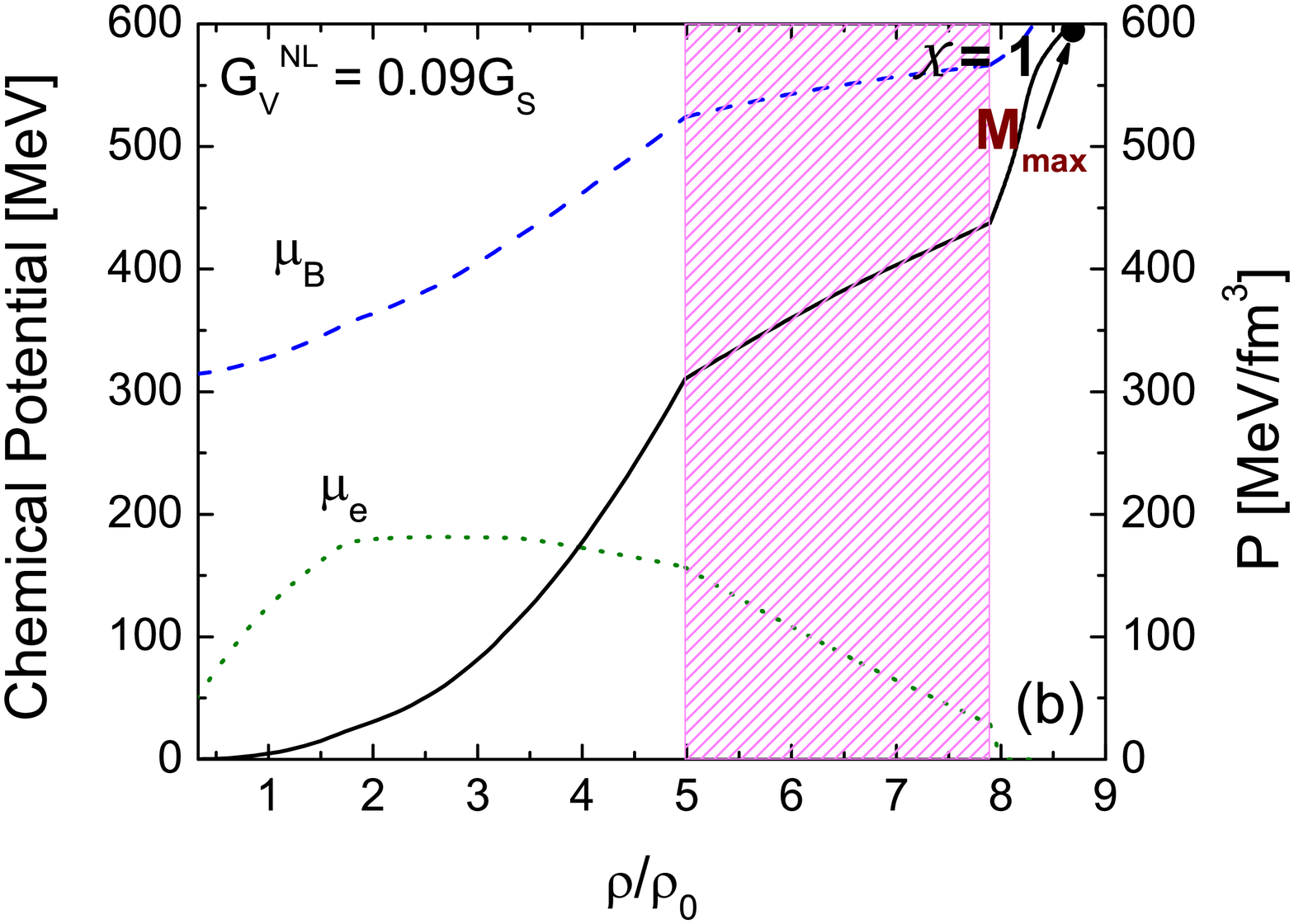}
\caption{(Color online) Pressure $P$ (solid lines), baryon chemical
potential $\mu_B = \mu_n/3$ (dashed lines), and electron chemical
potential $\mu_e$ (dotted lines) as a function of baryon number
density (in units of $\rho_0 = 0.16 \, {\rm fm}^{-3}$) for
parametrization GM1. The hatched areas denote mixed phase regions
where confined hadronic matter and deconfined quark matter
coexist. The data in panels (a) and (b) is for vector interactions
$0.075 G_S$ and $0.09 G_S$, respectively.}
\label{gm1press}
\end{figure}

Since we model the quark-hadron phase transition in three-space,
accounting for the fact that the electric and baryonic charge are
conserved for neutron star matter, the pressure varies monotonically
with the proportion of the phases in equilibrium, as shown in
Figs.\ \ref{fig:eos_3D0} -- \ref{fig:eos_3D} and \ref{press} --
\ref{gm1press}. For the latter, the hatched areas denote the mixed
phase regions where
\begin{table}[htb]
\begin{center}
\caption{\label{tabla2} Widths of mixed phases and central densities
  of maximum-mass neutron stars for the local NJL model of this paper
  ($\rho_0 = 0.16 \, {\rm fm}^{-3}$).}
\begin{ruledtabular}
\begin{tabular}{rcc}
$G_V^{L}/G_S$  ~~~&~~~Mixed phase    ~~~&~~~Central density of $M_{\rm max}$ \\
          &($\rho_0$)              &($\rho_0$) \\
\hline
$0$ (GM1)  & $3.64-7.60$ &  $7.21$ \\
$0.30$ (GM1) &  $5.07-8.92$ & $8.81$  \\
$0$ (NL3)  & $2.85-4.51$ &  $5.96$ \\
$0.30$ (NL3) &  $3.33-6.03$ & $6.52$  \\
\end{tabular}
\end{ruledtabular}
\end{center}
\end{table}
confined hadronic matter and deconfined quark matter coexist.  The
quark matter contents of the maximum-mass neutron stars computed for
these equations of state are indicated ($\chi$ values).  Pure quark
matter exists in stars marked with $\chi=1$.  Our calculations show
that, in the non-local case, similarly to the observation in
Ref.\ \cite{Orsaria2013}, the inclusion of the quark vector coupling
contribution shifts the onset of the phase transition
\begin{table}[htb]
\begin{center}
\caption{\label{tabla3} Widths of mixed phases and central densities
  of maximum-mass neutron stars for the non-local NJL model of this
  paper ($\rho_0 = 0.16 \, {\rm fm}^{-3}$).}
\begin{ruledtabular}
\begin{tabular}{rcc}
$G_V^{NL}/G_S$ ~~~&~~~Mixed phase  ~~~&~~~Central density of $M_{\rm max}$\\
           &($\rho_0$)  &($\rho_0$)\\ \hline
$0$ (GM1) & $3.22-8.18$ & $6.87$    \\
$0.09$ (GM1)&  $4.98-7.90$ & $8.69$  \\
$0$ (NL3) & $2.71-6.87$ & $5.68$    \\
$0.09$ (NL3)&  $3.24-6.31$ & $6.28$  \\
\end{tabular}
\end{ruledtabular}
\end{center}
\end{table}
to higher densities and narrows the width of the mixed quark-hadron
phase, when compared to the case $G_V = 0$. To the contrary, when the
quark matter phase is represented by the local l3NJL model, the width
of the mixed phase tends to be broader for finite $G_V$ values.
This effect can be seen both in Fig.\ \ref{press} as well as in Tables
\ref{tabla2} and \ref{tabla3}.

To account for the uncertainty in the theoretical predictions of the
ratio $G_V/G_S$
\cite{Sasaki:2006ws,Fukushima:2008wg,Bratovic:2012qs,Contrera:2012wj},
we treat the vector coupling constant as a free parameter.  We
observed that the n3NJL model is more sensitive to the increase of
$G_V/G_S$ than the local model. For $G_V/G_S\,>\,0.09$ we have a shift
of the onset of the quark-hadron phase transition to higher and higher
densities, preventing quark deconfinement in the cores of neutron
stars.  However, we can reach values of $G_V/G_S$ up to $0.3$ for the
local NJL case.

\begin{figure*}[!htbp]
\centering
\hspace*{\fill}%
       \subfigure[Mass-radius relationships of neutron stars made of
          quark-hybrid matter.]{%
            \label{fig:firstl}%
            \includegraphics[width=0.43\textwidth]{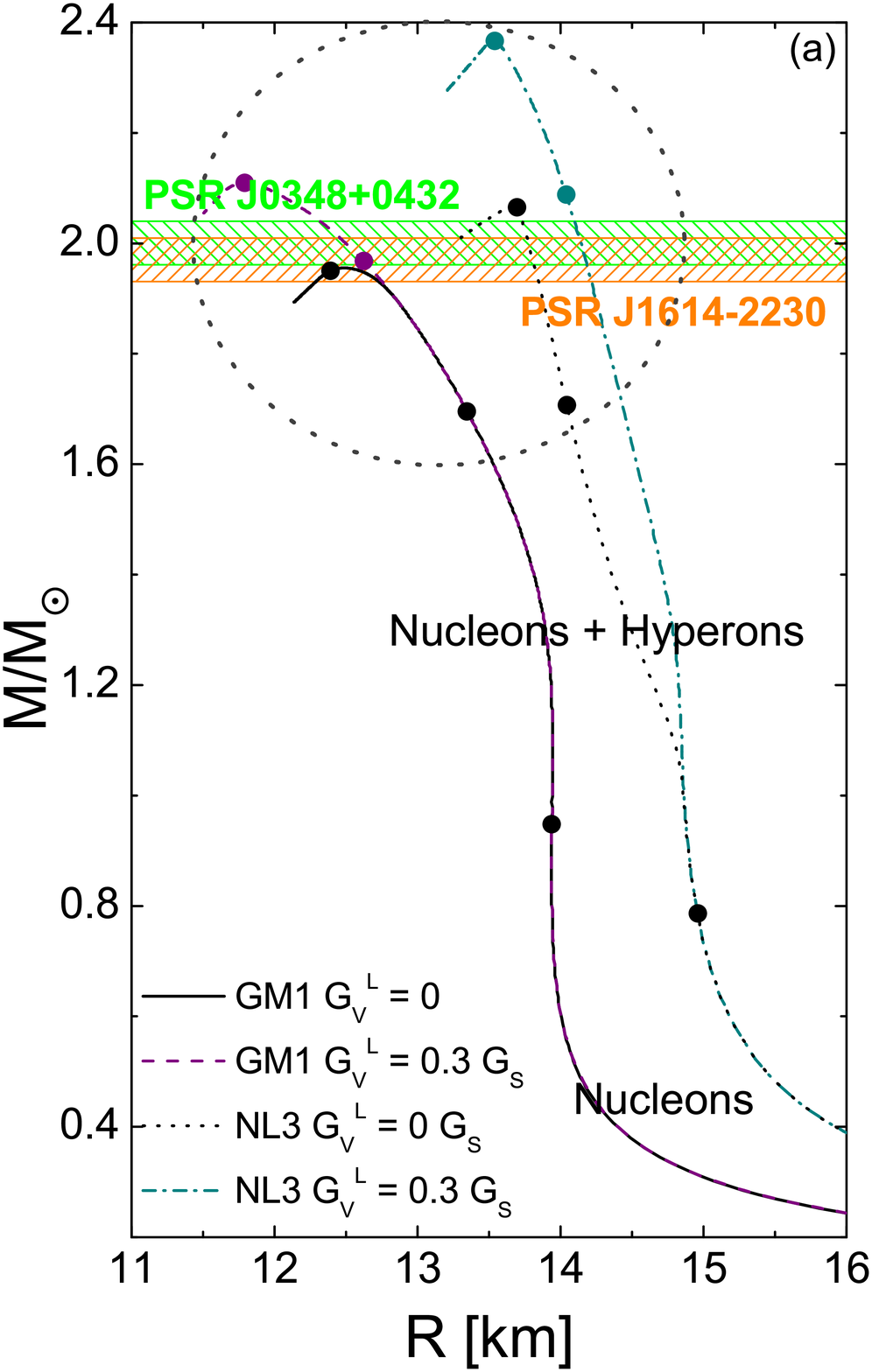}} \hfill%
       \subfigure[Enlargement of the circled region of
       Fig.\ \ref{fig:firstl}.]{%
           \label{fig:secondl} %
           \includegraphics[width=0.45\textwidth]{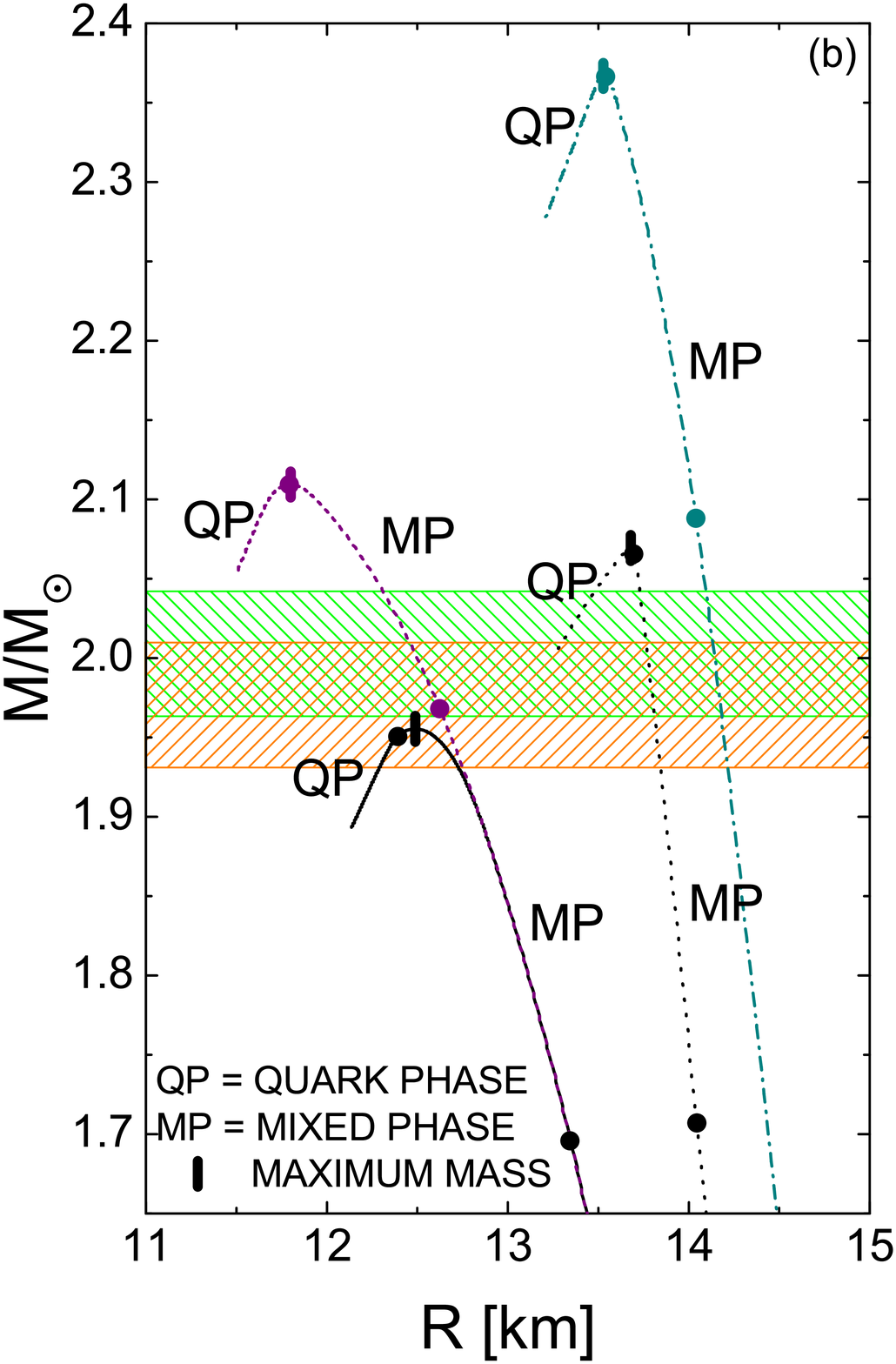}} %
\hspace*{\fill}%
\caption{(Color online) Quark-hybrid matter inside of neutron stars,
  computed for the local NJL model (l3NJL) and hadronic model
  parametrizations GM1 and NL3. The symbols 'MP' and 'QP' stand for
  mixed phase and pure quark phase, respectively. The vertical bars
  denote the maximum-mass star of each stellar sequence.}
  \label{mrl}
\end{figure*}

\begin{figure*}[!htbp]
\centering
\hspace*{\fill}%
       \subfigure[Mass-radius relationships of neutron stars made of
       quark-hybrid matter.]{%
            \label{fig:firstnl}%
            \includegraphics[width=0.45\textwidth]{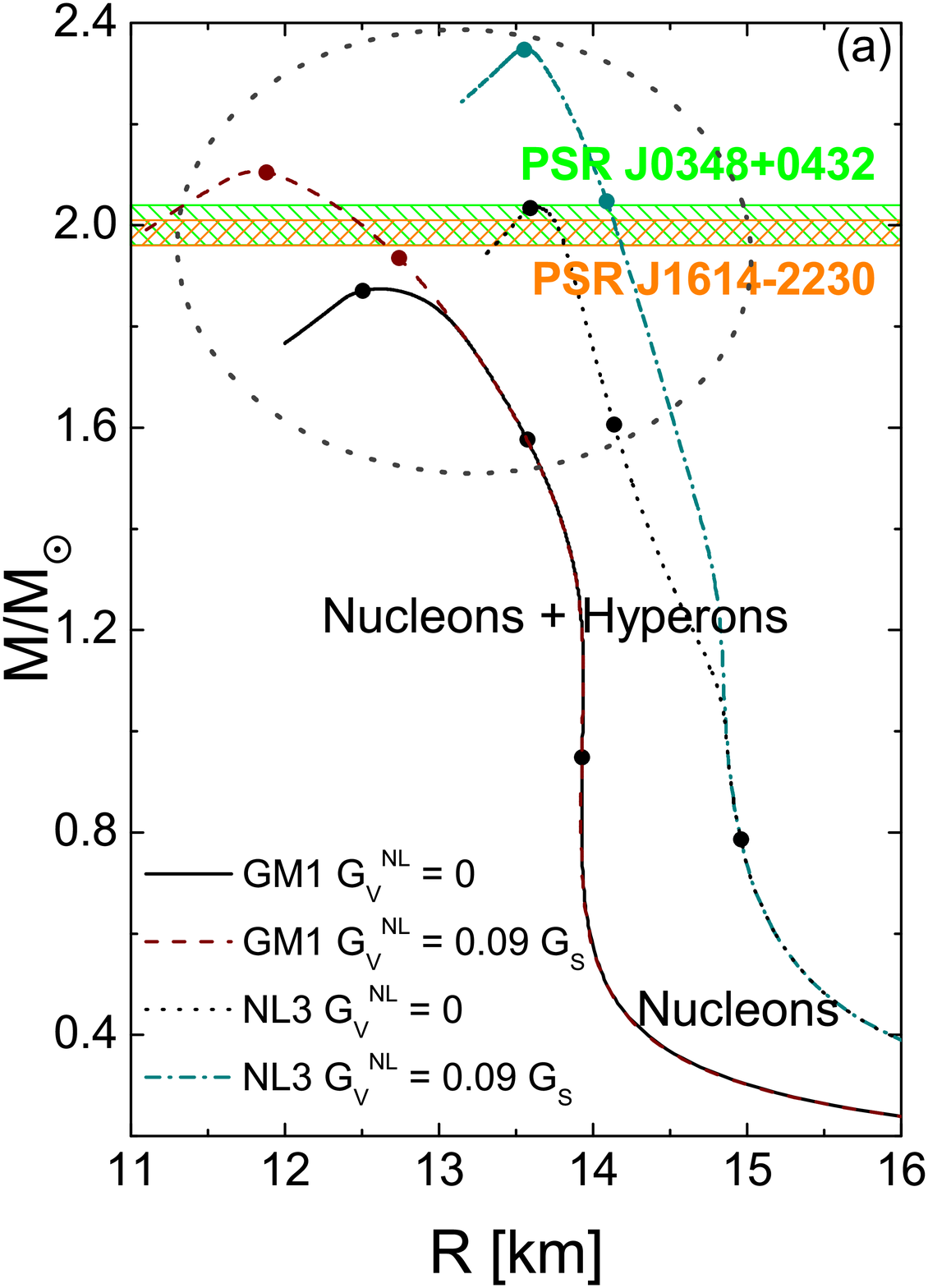}} \hfill%
       \subfigure[Enlargement of the circled region of
       Fig.\ \ref{fig:firstnl}.]{%
           \label{fig:secondnl} %
           \includegraphics[width=0.44\textwidth]{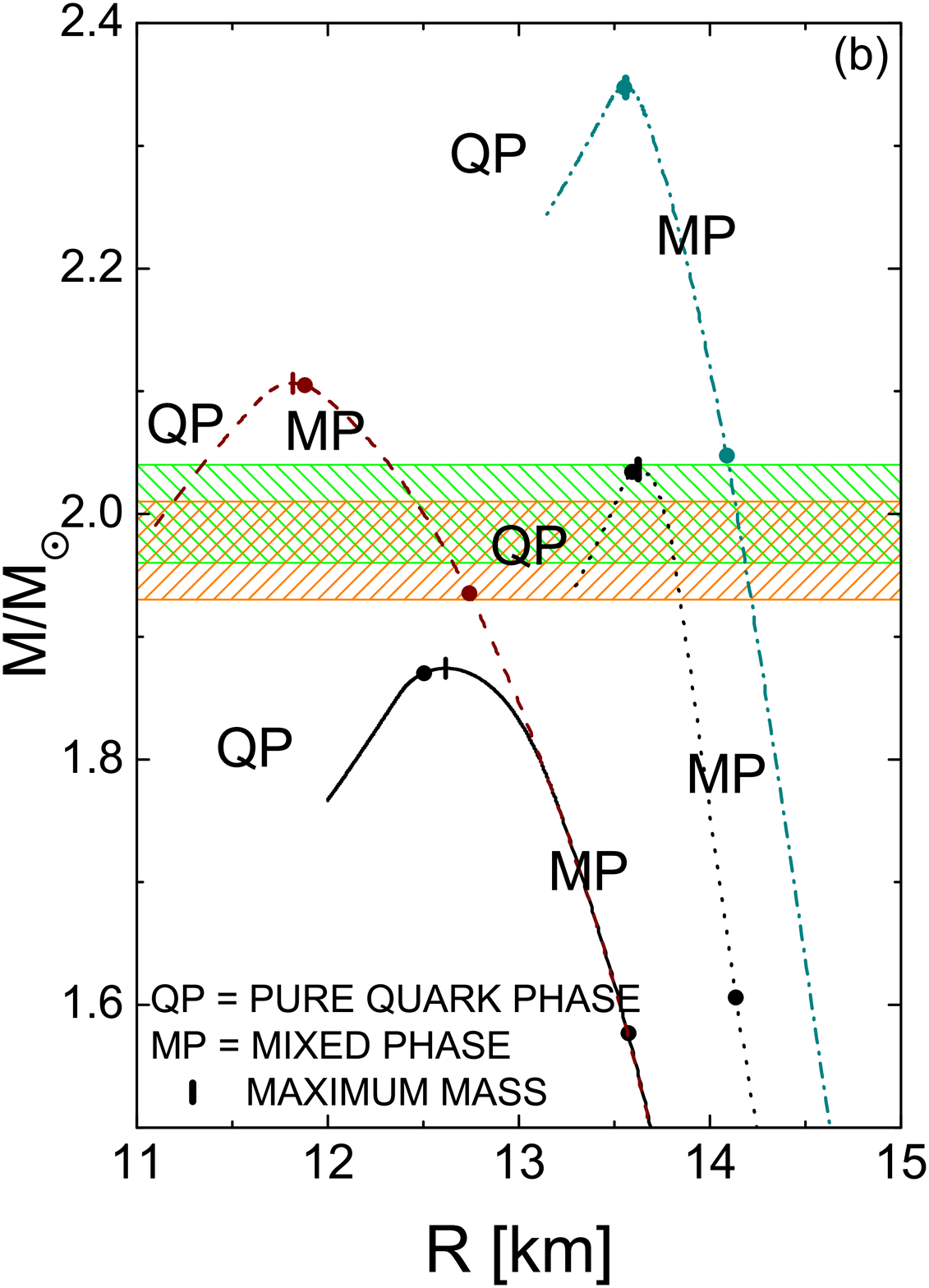}} %
\hspace*{\fill}%
\caption{(Color online) Same as Fig.\ \ref{mrl}, but computed for the
  non-local NJL model (n3NJL) and hadronic model parametrizations GM1 and NL3.}
 \label{mrnl}
\end{figure*}

Next we determine the bulk properties of spherically symmetric neutron
stars for the collection of equations of state discussed in this
paper.  The properties are computed from the
Tolmann-Oppenheimer-Volkoff (TOV) equation of general relativity
theory \cite{glendenning00:book,fridolin1,Tolman39}.  The outcomes are
shown in Figs.\ \ref{mrl} and \ref{mrnl} for the local and non-local
NJL model, respectively.  The green and orange bands in
Fig.\ \ref{mrl} display the masses of the recently discovered, massive
neutron stars PSR J1614-2230 \cite{Demorest2010} and J0348+0432
\cite{Antoniadis13}, respectively.

 As a neutron star becomes increasingly more massive, the stellar core
 composition consist of either only nucleons, nucleons and hyperons, a
 mixed phase (MP) of quarks and hadrons, or a pure quark matter phase
 (QP), as indicated in Figs.\ \ref{mrl} and \ref{mrnl}.  For
 the non-local NJL model and the NL3 parametrization for the hadronic
 phase (Fig.\ \ref{fig:secondnl}), we find that pure quark does not
 exist in stable neutron stars.  Only neutron stars which lie to the
 left of the mass peak are dense enough to contain such matter. These
 stars are however unstable against radial oscillations and thus
 cannot exit stably in the universe. The situation is different if the
 GM1 parameter set is used for the hadronic lagrangian. In this case a
 small amount of pure quark matter turns out to exist in the
 maximum-mass neutron star if the strength of the vector repulsion is non-zero
 ($G_V^{NL}/G_S = 0.09$). (See also Table \ref{tabla3}.) Extended regions of a
mixed phases of quarks and hadrons are found for both hadronic
parametrizations.
\begin{table}[htb]
\begin{center}
\caption{Maximum masses and radii of neutron stars made of
  quark-hybrid matter for different vector repulsion ($G_V/G_S$).}
\label{tabla4}
\begin{ruledtabular}
\begin{tabular}{cccc}
\hspace*{.1cm} NJL Model \hspace*{.1cm} & \hspace*{.1cm}
$G_V/G_S$ \hspace*{.1cm} & \hspace*{.1cm} $R_{\rm max}$
(km) \hspace*{.1cm} & \hspace*{.1cm}
$M_{\rm max}/M_{\odot}$ \hspace*{.1cm} \\ \hline
 Local &  0    & 12.49   &   1.96  \\
 GM1   &  0.30 & 11.80   &   2.11 \\
\hline
 Local &  0    & 13.68   &   2.07  \\
  NL3  &  0.30 & 13.53   &   2.37 \\
\hline
Non-local&  0    &  12.62    &    1.87 \\
   GM1  &   0.09 &  11.81 &    2.11 \\
\hline
Non-local&   0 &  13.62 &    2.04  \\
    NL3 &   0.09 &  13.56 &    2.35  \\
\end{tabular}
\end{ruledtabular}
\end{center}
\end{table}
The situation is somewhat different for the local NJL model
(Fig.\ \ref{mrl}), for which pure quark matter cores are obtained for
NL3 (see Table \ref{tabla2} and Fig.\ \ref{fig:secondl}). The results
for the masses and radii of the maximum-mass neutron stars are shown
in Table \ref{tabla4}.

\section{Summary and conclusions}
\label{sect:7}

 In this work we use extensions of the local (l3NJL) and non-local
(n3NJL) Nambu-Jona Lasinio model to analyze the possible occurrence of
quark deconfinement in the cores of neutron stars. We have constructed
the phase transition from hadronic matter to quark matter via the
Gibbs conditions, imposing global electric charge neutrality and
baryon number conservation. We have calculated the mass-radius
relationships of ordinary neutron stars and neutron stars with
deconfined quarks in their centers (quark-hybrid stars). Depending on
the strength of the quark vector repulsion, we find that mixed phases of
confined hadronic matter and deconfined quarks can exist in neutron
stars as massive as  $2.1$ to $2.4\, M_\odot$. The radii of these
objects are between 12 and 13~km, as expected for neutron stars.

According to our study, for the n3NJL model, a transition to pure
quark matter occurs only in neutron stars which lie on the
gravitationally unstable branch of the stellar sequences if the
parametrization NL3 for the hadronic matter lagrangian is used. This
is different for the other hadronic parametrization considered in this
paper, GM1, which predicts, for sufficiently large vector interactions
among the quarks, pure quark matter cores in maximum-mass neutron
stars. Pure quark matter cores are also obtained for the l3NJL model
if the NL3 parametrization is being use.  For the GM1 parametrization,
however, pure quark matter is not present but gives way to an extended
region of deconfined quarks which are in chemical equilibrium with
various baryon species.

The latter is found to exist in all neutron stars models, independent
of the value of the vector repulsion among quarks.  With increasing
stellar mass, all the stellar core compositions consist either of
nucleons only, nucleons and hyperons, a mixed phase of quarks and
hadrons (MP), and, in some cases, a pure quark matter phase (QP) in
the core.

\section*{Acknowledgments}
M.\ Orsaria thanks N.\ N.\ Scoccola and T. Hell for helpful
discussions.  M.\ Orsaria and G.\ Contrera thank CONICET for financial
support. H.\ Rodrigues thanks CAPES for financial support under
contract number BEX 6379/10-9. F.\ Weber acknowledges supported by the
National Science Foundation (USA) under Grant PHY-0854699.

\end{document}